\documentclass[prd,superscriptaddress,amsfonts,amssymb,amsmath,showpacs]{revtex4-2}
\usepackage{bm}
\usepackage{amsfonts}
\usepackage{latexsym}
\usepackage[latin1]{inputenc}
\usepackage{graphicx}
\usepackage{amsmath}
\usepackage{textcomp}
\usepackage{float}
\usepackage{booktabs}
\usepackage{dcolumn}
\usepackage{booktabs}
\usepackage{multirow}
\usepackage{hyperref}
\hypersetup{colorlinks,citecolor=blue}
\usepackage{amsmath}
\usepackage{xcolor}
\usepackage{orcidlink}
\usepackage[caption=false]{subfig}
\usepackage{commath}

\def\jnl@style{\it}
\def\aaref@jnl#1{{\jnl@style#1}}

\def\aaref@jnl#1{{\jnl@style#1}}

\def\aj{\aaref@jnl{AJ}}                   
\def\apj{\aaref@jnl{ApJ}}                 
\def\apjl{\aaref@jnl{ApJ}}                
\def\apjs{\aaref@jnl{ApJS}}               
\def\apss{\aaref@jnl{Ap\&SS}}             
\def\aap{\aaref@jnl{A\&A}}                
\def\aapr{\aaref@jnl{A\&A~Rev.}}          
\def\aaps{\aaref@jnl{A\&AS}}              
\def\mnras{\aaref@jnl{Mon.~Not.~Roy.~Astron.~Soc.}}             
\def\prd{\aaref@jnl{Phys.~Rev.~D}}        
\def\prc{\aaref@jnl{Phys.~Rev.~C}}  
\def\prl{\aaref@jnl{Phys.~Rev.~Lett.}}    
\def\qjras{\aaref@jnl{QJRAS}}             
\def\skytel{\aaref@jnl{S\&T}}             
\def\ssr{\aaref@jnl{Space~Sci.~Rev.}}     
\def\zap{\aaref@jnl{ZAp}}                 
\def\nat{\aaref@jnl{Nature}}              
\def\aplett{\aaref@jnl{Astrophys.~Lett.}} 
\def\apspr{\aaref@jnl{Astrophys.~Space~Phys.~Res.}} 
\def\physrep{\aaref@jnl{Phys.~Rep.}}      
\def\physscr{\aaref@jnl{Phys.~Scr}}       
\def\commat{\aaref@jnl{Comm.~Math.~Phys.}}              
\def\science{\aaref@jnl{Science}}               
\def\cqg{\aaref@jnl{Classical Quant.~Grav.}}            
\def\jpcs{\aaref@jnl{JPCS}}                                     
\def\ijmpd{\aaref@jnl{Int.~J.~Mod.~Phys.~D}}                    
\def\grg{\aaref@jnl{Gen.~Relat.~Gravit.}}               
\def\rpp{\aaref@jnl{Rep.~Prog.~Phys.}}          
\def\npa{\aaref@jnl{Nucl.~Phys.~A}}        
\def\lrr{\aaref@jnl{Living Rev.~Rel.}}                   
\def\jcap{\aaref@jnl{J.~Cosmology Astropart.~Phys.}}    
\def\rmp{\aaref@jnl{Rev.~Mod.~Phys.}}   
\def\epjc{\aaref@jnl{Eur.~Phys.~J.~C}}

\allowdisplaybreaks[1]
\renewcommand{\arraystretch}{1.1}
\begin{document}

\color{black}       
\title{\bf Quintessence model in the late Universe} 
\author{L.K. Duchaniya\orcidlink{0000-0001-6457-2225}}
\email{duchaniya98@gmail.com}
\affiliation{Wilczek Quantum Center, Shanghai Institute for Advanced Studies, Shanghai 201315  \\University of Science and Technology of China, Hefei 230026, China}

\author{Jackson Levi Said\orcidlink{0000-0002-7835-4365}}
\email{jackson.said@um.edu.mt}
\affiliation{Institute of Space Sciences and Astronomy, University of Malta, Malta, MSD 2080}
\affiliation{Department of Physics, University of Malta, Malta}

\author{B. Mishra\orcidlink{0000-0001-5527-3565} }
\email{bivu@hyderabad.bits-pilani.ac.in }
\affiliation{Department of Mathematics, Birla Institute of Technology and Science, Pilani, Hyderabad Campus, Jawahar Nagar, Kapra Mandal, Medchal District, Telangana 500078, India.}

\author{Yu Shi\orcidlink{0000-0002-2910-0684}}
\email{yu\_shi@ustc.edu.cn}
\affiliation{Wilczek Quantum Center, Shanghai Institute for Advanced Studies, Shanghai 201315  \\University of Science and Technology of China, Hefei 230026, China}

\begin{abstract}
Scalar fields have shown great potential in inducing tailored modifications compared to cosmic evolution in the $\Lambda$CDM model. We investigate a quintessence model with five physically motivated scalar field potentials: the general power-law, quadratic, fifth-power, hyperbolic sinh, and axion-like potentials. The model parameters are constrained by using the late-time cosmological observations, including Cosmic Chronometers (CC), Pantheon$^{+}$, BAO, and DESI DR2 data. To assess the influence of the local distance-ladder calibration, we additionally perform a separate analysis using the Pantheon$^{+}$\& SH0ES compilation. Our analysis shows that the inclusion of the SH0ES calibration systematically shifts the inferred Hubble constant toward higher values, whereas the addition of BAO and DESI DR2 data prefers comparatively lower values and significantly reduces parameter degeneracies. We center the action of these models in the late Universe which leaves early $\Lambda$CDM cosmology unchanged. Across all five potentials, an inverse correlation between $H_0$ and $\Omega_{m0}$ is consistently observed, indicating that higher values of the Hubble constant are associated with a lower present-day matter density and a stronger contribution from dark energy to the current accelerated expansion.
\end{abstract}

\maketitle

\section{Introduction} \label{SEC_intro}

The wealth of evidence gathered over the last few decades on  both  astrophysical and cosmological scales  supports  the standard model of cosmology \cite{Misner:1974qy,Clifton:2011jh}, in which cold dark matter (CDM) acts as a stabilizing agent in galactic scales systems \cite{Baudis:2016qwx,Bertone:2004pz} while dark energy appears through a cosmological constant $\Lambda$ \cite{Peebles:2002gy,Copeland:2006wr}. Despite these successes, the prospect of directly measuring dark matter particles remains elusive \cite{Gaitskell:2004gd}, while internal consistency issues persist in the cosmological constant realization of dark energy \cite{Weinberg:1988cp}. Recently, other issues have arisen with the standard model of cosmology. These problems first appeared in the form of statistically significant differences in the expansion rate of the Universe, in the value of the Hubble constant \cite{DiValentino:2020zio}. One perspective is that cosmology-independent local, or late-time, measurements of the Hubble constant $H_0$ \cite{Riess:2019cxk,Shajib:2023uig}, and measurements based on a $\Lambda$CDM using global, or early time, survey data \cite{Aghanim:2018eyx,Ade:2015xua}, are showing a discrepancy in the fundamental physics associated with describing cosmic evolution. While the issue may be related in some part to systematics, other measurements \cite{Baker:2019nia,2017arXiv170200786A,Barack:2018yly} may give further information on the nature of the possible new physics that may describe the breadth of data available \cite{Abdalla:2022yfr}. Along a similar vein, there are also growing questions on whether this tension has permeated into the description of the large-scale structure and its evolution \cite{Abdalla:2022yfr,Benisty:2020kdt,Poulin:2022sgp,DiValentino:2020vvd,LeviSaid:2021yat, divalentino2025cosmoversewhitepaperaddressing}.

The cosmic microwave background (CMB) data reflect the state of the Universe at redshift z$\approx$1090, merely 400,000 years following the Big Bang. Baryon acoustic oscillations (BAO) derived from galaxy surveys limit the expansion history of the Universe within the range of $0.1<z<4.2$, encompassing both the matter-dominated phase and the recent period of cosmic acceleration. Type Ia supernovae (SNe Ia) observations were utilized to uncover dark energy and to refine the understanding of the expansion history at lower redshifts. The collective insights from these cosmological probes throughout the evolving Universe offer quantitative data on the temporal transformation of dark energy density. Recently, the DESI collaboration unveiled a cosmological analysis based on the most current BAO measurements from its Data Release 2 (DR2) \cite{Abdul_Karim_2025_Desi}. This analysis indicates that discrepancies among datasets are becoming increasingly significant within the $\Lambda$CDM framework, suggesting the idea of evolving dark energy as a potential resolution and also providing updated limits on the neutrino mass \cite{elbers2025constraintsneutrinophysicsdesi}. Specifically, while the DESI DR2 BAO results align broadly with the $\Lambda$CDM cosmological model, they reveal a discrepancy of 2.3$\sigma$ \cite{Abdul_Karim_2025_Desi} when compared with Planck CMB data. Conversely, a time-evolving dark energy component is supported by the combined analysis of these datasets, a situation where the datasets show consistency. This pattern can be inferred through simple CPL parametrizations \cite{Linder_2003_para} or other techniques for reconstructing dark energy \cite{lodha2025extendeddarkenergyanalysis}. Employing the $\omega_0 \omega_a$CDM framework, the combination of DESI BAO DR2 with CMB temperature and polarization fluctuations, along with CMB lensing, suggests that evolving dark energy is favored at a significance level of $3.1\sigma$. This preference varies between $2.8\sigma$ and $4.2\sigma$ when SNe Ia data are included, depending on the specific SNe sample utilized \cite{Abdul_Karim_2025_Desi} (this preference is also corroborated by the Dark Energy Survey's combined analysis of BAO and SNe \cite{descollaboration2026darkenergysurveyimplications}).

It is increasingly becoming less likely that the series of cosmological tensions is the result of a single systematic issue with the statistical treatment, there have been a diversity of possibilities of directions beyond $\Lambda$CDM in terms of additional or new physical mechanisms. In the latter case, there have been several promising proposals including the modification of physics beyond recombination such as in early dark energy \cite{Poulin:2023lkg}, the additional of extra relativistic degrees of freedom \cite{DiValentino:2021imh}, as well as the modification of gravitational physics \cite{Addazi:2021xuf,CANTATA:2021ktz,Bahamonde:2021gfp,Bamba:2012cp, Nojiri:2010wj,Nojiri:2017ncd} at various scales. In these scenarios, scalar-tensor theories \cite{Joyce:2016vqv,Palti:2019pca,Capozziello:2007ec, Gonzalez-Espinoza:2020jss} have been prominent as providing a possible avenue for confronting the problem of cosmic tensions. The simplest of these models involves a canonical scalar field as an additional ingredient to $\Lambda$CDM. In these scenarios, the cosmological constant is supplanted by the scalar field. This is analogous to the action of scalar fields in inflation theory \cite{Guth:1980zm,Linde:1981mu, Gonzalez_Espinoza_2020A}. 

Both canonical and nonminimally coupled scalar-tensor theories can be collectively studied through Horndeski gravity \cite{Horndeski:1974wa,Horndeski:2024sjk,Kobayashi:2019hrl} where a single scalar field is used to construct the most general framework in which second order equations of motion are produced. While complex, this framework can produce concise expressions for different phenomenology. On the other hand, recent multimessenger constraints from the gravitational wave sector have put severe constraints on the most exotic elements of this formulation \cite{LIGOScientific:2017zic,Ezquiaga:2017ekz}. While alternative geometric formulations \cite{Bahamonde:2019shr,Bahamonde:2022cmz}, and beyond Horndeski gravity \cite{Gleyzes:2014dya,Traykova:2019oyx} proposals have yielded interesting results in this sector, it may also be the case that a simple canonical scalar field should be reassessed. These scalar field theories give very promising results when their action is considered in the early Universe \cite{DiValentino:2020vvd,Poulin:2022sgp,Kazantzidis:2018rnb,Gonzalez-Espinoza:2021mwr,Asghari:2019qld,Jimenez:2024lmm,Figueruelo:2021elm,BeltranJimenez:2021wbq,BeltranJimenez:2022irm,BeltranJimenez:2024lml}.

In this work, we reconsider the action of a single scalar field in the local Universe through the realization of different potentials for a canonical scalar field. This is performed by considering the Friedmann and Klein-Gordon equations in Sec.~\ref{SEC_matheforma} together with a series of late time data sets in Sec.~\ref{SEC_observational_data}. A Markov chain Monte Carlo (MCMC) approach is taken in Sec.~\ref{cosmologicalmodels} where five potentials are considered for the scalar field. The potentials we are considering include a regular power-law form, a quadratic potential, and a fifth power potential. We also explore a hyperbolic form, which is inspired by its behavior resembling a tracker-like solution under certain conditions, as well as an axion-like potential. By examining these potentials, we aim to reassess the capabilities of scalar field cosmologies to address some of the observational challenges that have emerged in recent years. In Sec.~\ref{modelcomparison} we compare these models in the context of $\Lambda$CDM, and then finally close with a summary in Sec.~\ref{conclusion}.

\section{Quintessence cosmology}\label{SEC_matheforma}

The global cartesian coordinate system aligned with the Friedman-Robertson-Walker (FRW) cosmological model defines the spacetime interval between two events as follows
\begin{equation}\label{metric}
ds^{2} = -N(t)^2 dt^{2}+a(t)^2 (dx^2+dy^2+dz^2)\,,   
\end{equation} 
where $a(t)$ define the scale factor and $N(t)$ stands for lapse function. We have taken the following quintessence model \cite{Bartolo_1999, Faraoni_2000, Geng:2011aj, Geng:2011ka}
\begin{equation}\label{action_formula}
S = \int d^4x \, \sqrt{-g} \left(\frac{R}{16\pi G} - \frac{1}{2}\,g^{\mu\nu} \partial_\mu \phi \, \partial_\nu \phi - V(\phi) \right) + S_m + S_r\,.  
\end{equation}

Where \( g \) is the determinant of the metric \( g_{\mu\nu} \), \( R \) is the Ricci scalar, \( S_m \) denotes the matter Lagrangian, and \( S_r \) denotes the radiation Lagrangian. If we vary the action \eqref{action_formula} with respect to scalar factor $a(t)$ and lapse function $N(t)$, we get the following motion equations,
\begin{eqnarray}
 3H^{2} &=& 8 \pi G \left(\frac{\dot{\phi}^2}{2}+ V(\phi)+\rho_m+\rho_r\right) \,, \label{model_first_friedmann_equation}\\
 2 \dot{H}&=&-8 \pi G \left(\dot{\phi}^2+\rho_m+\frac{4}{3}\rho_r\right)\,.\label{model_second_friedmann_equation}
\end{eqnarray}

Where $H=\frac{\dot{a}}{a}$ is the Hubble parameter and over dot defines the derivative for cosmic time $t$. $\rho_{m}$ and $\rho_{r}$ stand for the matter and radiation energy density respectively. In this context, $\frac{\dot{\phi}^2}{2}$ denotes the kinetic term, while $V(\phi)$ represents the potential function. Furthermore, by varying \eqref{action_formula} for scalar field $\phi$, we have obtained the following the equation
\begin{equation}\label{Klein_Gordan_equation}
\Ddot{\phi}+ 3 H \dot{\phi} + V^{'}(\phi)=0\,.  \end{equation}
Where prime (') denotes the derivative to scalar field $\phi$. In this work, we will study various cosmological observation data sets, including $H_0$ prior, in the context of the quintessence model. Formalisms for the data sets are described in the section-\ref{SEC_observational_data}. To do so, we need to calculate the Hubble parameter $H(z)$ in terms of redshift $z$ with the help of Eqs.~( \ref{model_first_friedmann_equation},\ref{Klein_Gordan_equation}). Additionally, we need to specify the particular form of the potential function $V(\phi)$. Therefore, we will use five different potential function forms and constrain their parameters using observation data. As a function of redshift $z$, the Klein-Gordan Eq.~ \eqref{Klein_Gordan_equation}  can be written as 
\begin{equation}\label{redshift_Klein_gordan_equation}
(1+z)^2 H^2(z) \frac{d^2\phi}{dz^2}+ (1+z)^2 H(z) \frac{dH}{dz}\frac{d\phi}{dz}-2(1+z)H^2(z)\frac{d\phi}{dz}+V_{,\phi}(\phi) \frac{d\phi}{dz}=0 \,, 
\end{equation}

where $V_{,\phi}(\phi)$ represents the derivative with respect to scalar field $\phi$. The Eq.~ \eqref{model_first_friedmann_equation} can be written as a function of redshift $z$ as,
\begin{equation}\label{model_Hz_first_friedmann_equation}
H^2(z)=\frac{3H_0^2(\Omega_{m0} (1+z)^3+\Omega_{r0} (1+z)^4)+8\pi G V(\phi)}{3-4 \pi G (1+z)^2 \left(\frac{d\phi}{dz}\right)^2}\,.
\end{equation}
In the above equation, $H_0$ refers to the Hubble parameter value, $\Omega_{m0}$, and $\Omega_{r0}$ refers to the matter density parameter and the radiation density parameter, respectively, at present.



 

\section{ Observational Data}\label{SEC_observational_data}

The ensuing work will concentrate on data sets obtained through observational studies. This study used the Cosmic Chronometers (CC), Pantheon+ SH0ES [PN$^+$\&SH0ES], Pantheon+ [PN$^{+}$], BAO and DESI DR2 data set. We used the publicly available \texttt{emcee package}, found at Ref. \cite{Foreman_Mackey_2013}, to conduct an MCMC analysis for the combination of the data sets and five different potential functions $V(\phi)$. MCMC is a type of sampling technique that generates samples from the posterior distribution of the model. This technique is often used in Bayesian statistics, where the posterior distribution estimates model parameters. MCMC can also be used to estimate the uncertainty of a model. This process requires a large sample size to ensure the generated models represent the data. Thus, for every parameter, we derive both its one and two dimensional distributions; the one-dimensional distribution shows the posterior distribution of the parameters, while the two-dimensional distribution shows the covariance between two distinct parameters. For this, we have performed the 1$\sigma$ (68\%) and 2$\sigma$ (95\%) confidence regions are generated using the \texttt{ChainConsumer package} \cite{Hinton2016}. One of current cosmology's most significant open questions is the discrepancy between the Hubble constant \( H_0 \) from early-Universe probes and those obtained from late-time observations. Scalar field theories extend Einstein's general relativity by introducing a dynamic scalar field that interacts with gravity and represents a well-founded and adaptable class of models for investigating such extensions. These theories can seamlessly incorporate time-dependent changes to the effective gravitational constant and permit evolving behaviors of dark energy that can impact the late-time expansion history of the Universe. Specifically, the scalar field can be configured to remain insignificant during the early Universe, thereby maintaining the effectiveness of \( \Lambda \)CDM in explaining CMB and structure formation while becoming dynamically important in recent cosmic history. This gravity modification at late times may result in an accelerated expansion rate in the nearby Universe, effectively raising the inferred value of \( H_0 \) and possibly aligning it with local measurements.

Additionally, scalar field theories commonly appear as low-energy limits of higher-dimensional or string-inspired frameworks, which adds theoretical validity to their exploration. Their adaptability also facilitates the inclusion of screening mechanisms that ensure alignment with local gravity tests. Consequently, examining scalar-tensor theories in late-time cosmology is both timely and necessary, as it presents a potential solution to the \( H_0 \) tension while aligning with a wide array of cosmological and astrophysical observations. In this regard, exploring the dynamic influence of scalar fields on recent expansion history could yield important insights into the characteristics of dark energy and the fundamental principles governing our Universe.



\textbf{Cosmic Chronometers (CC):} We utilized 31 data points, estimated using the CC method, as reported in \cite{Zhang_2014hz, Jimenez_2003cmb, Moresco_2016hubb, Simon_2005prd, M_Moresco_2012JCAP, Daniel_Stern_2010jcap, Moresco_2015mnras}. Using this approach, we may directly understand the Hubble function at various redshifts, up to $ z \lesssim 2 $. Since CC data relies on measuring the age difference between two passively evolving galaxies that formed simultaneously but are separated by a small redshift interval ($\frac{\Delta z}{\Delta t}$), it is more effective than other methods that depend on determining the absolute age of galaxies \cite{Jimenez_2002}. The Hubble data set is linked to star ages, which are derived from robust stellar population synthesis models \cite{G_mez_Valent_2018jcap, L_pez_Corredoira_2017aa}, even though they do not rest on a cosmological model or the Cepheid distance scale. The related estimate of $\chi^{2}_{CC}$ is given by
\begin{equation}\label{chisqure_hz}
\chi^2_{CC}= \Delta H(z_i, \Theta)^{T} \,C_{CC}^{-1} \,\Delta H(z_i, \Theta) \,,   
\end{equation}
where $\Delta H(z_i, \Theta)= H(z_i, \Theta)-H_{\text{obs}}(z_i)$ and $C_{CC}^{-1}$ is the covariance matrix given in \cite{Moresco_2020covariance}. The term $H(z_i, \Theta)$ represents the theoretical values of the Hubble parameter at a specific redshift $z_i$, while $H_{\text{obs}}(z_i)$ indicates the observed values of the Hubble parameter at $z_i$.

\textbf{Type Ia Supernovae data set :} The PN$^{+}\&$ SH0ES collection consists of 1,701 light curves \cite{Brout_2022panplus, Riess_2022panplus, Scolnic_2022panplus} derived from 1,550 spectroscopically verified Type Ia supernovae (SNe Ia). These data will be used to determine cosmological parameters as part of the Pantheon+ supernova study and the SH0ES distance-ladder analysis. The relative luminosity distance observations span the redshift range of $0.01 < z < 2.3$. The consistent intrinsic brightness of these supernovae renders them valuable for cosmological studies, as they enable us to determine distances to far-off galaxies by serving as standard candles. In particular, the distance modulus function is defined as the difference between the observed apparent magnitude \( m \) and its absolute magnitude \( M \). At redshift \( z_i \), the distance modulus function \( \mu(z_i, \Theta) \) can be expressed as,
\begin{equation}\label{modulus_function}
\mu(z_i, \Theta) = m - M = 5 \log_{10} \left[ D_L(z_i, \Theta) \right] + 25\,,    
\end{equation}
the luminosity distance $D_L(z_i, \Theta)$ can be written as 
\begin{equation}\label{luminosity_distance}
D_L(z_i, \Theta) = c(1 + z_i) \int_0^{z_i} \frac{dz'}{H(z', \Theta)}\,.  
\end{equation}

Where $c$ represents the speed of light, we may also marginalize $M$ as a nuisance parameter since each SNIa's apparent magnitude must be verified using a random fiducial absolute magnitude. The related estimate of $\chi^{2}_{SN}$ is given by \cite{Conley_2010Apjs}
\begin{equation}\label{chisquare_pantheon}
 \chi^2_{\text{SN}} = \left(\Delta\mu(z_i, \Theta)\right)^{T} C^{-1} \left(\Delta\mu(z_i, \Theta)\right)\,,
\end{equation}

where $C$ is the relevant covariance matrix that takes into account the systematic and statistical uncertainties, and $ \Delta\mu(z_i, \Theta) = \mu(z_i, \Theta) - \mu(z_i)_{\text{obs}}$.

\textbf{BAO data set :}  We also consider a joint data gathering of independent baryon acoustic oscillations (BAOs). The BAO data set includes observations from the six-degree field Galaxy Survey at an effective redshift of $z_{eff}$ = 0.106 \cite{Beutler_2011baosixdegree}, the BOSS DR11 quasar Lyman-alpha measurement at $z_{eff}$ = 2.4 \cite{du_Mas_des_Bourboux_2017}, and the SDSS Main Galaxy Sample at $z_{eff}$ = 0.15 \cite{Ross_2015}. Additionally, we consider the $H(z)$ measurements and the angular diameter distances from the SDSS-IV eBOSS DR14 quasar survey at $z_{eff} = \{0.98, 1.23, 1.52, 1.94\}$ \cite{Zhao_2018sdss_IV}, along with the consensus BAO measurements of the Hubble parameter and corresponding comoving angular diameter distances from the SDSS-III BOSS DR12 at $z_{eff}= \{0.38, 0.51, 0.61\}$ \cite{Alam_2017sdss_III}, where our MCMC analyses account for the complete covariance matrix in these two BAO data sets. We also analyze the BAO data from the DESI Data Release 2 (DR2) sample, which offers one of the most accurate geometric measurements of the late-time Universe. The DESI DR2 sample that encompasses low and intermediate redshifts from various galaxy tracers is detailed in \cite{Abdul_Karim_2025_Desi}. To study the BAO data set, it is necessary to define the Hubble distance $D_H(z)$, the comoving angular diameter distance $D_M(z)$, and the volume-average distance $D_V(z)$.
\begin{eqnarray}\label{BAO_distances}
D_H(z) = \frac{c}{H(z)},\quad D_M(z) = (1 + z)D_A(z), \quad D_V(z) = \left[(1 + z)^2 D_A^2(z) \frac{z}{H(z)}\right]^{1/3},
\end{eqnarray}

where the angular diameter distance is defined as \(D_A(z) = (1+z)^{-2} D_L(z)\). To utilize the reported BAO results for MCMC analyses, we need to take into account the relevant combination of parameters \( \mathcal{F}(z_i) = \bigg\{\frac{D_V(z_i)}{r_s(z_d)}, \frac{r_s(z_d)}{D_V(z_i)}, D_H(z_i),\\D_M(z_i)\bigg(\frac{r_{s,\text{fid}}(z_d)}{r_s(z_d)}\bigg), H(z_i)\bigg(\frac{r_s(z_d)}{r_{s, \text{fid}}(z_d)}\bigg), D_A(z_i)\bigg(\frac{r_{s, \text{fid}}(z_d)}{r_s(z_d)}\bigg)\bigg\} \). To compute this, we needed to find the comoving sound horizon \(r_s(z)\) at the redshift \(z_d \approx 1059.94\) \cite{Planck:2018vyg} after the baryon drag epoch.
\begin{eqnarray}\label{sound_horizon}
r_s(z) = \int_{z}^{\infty} \frac{c_s(\tilde{z})}{H(\tilde{z})} d\tilde{z} =\frac{1}{\sqrt{3}} \int_{0}^{1/(1+z)} \frac{da}{a^2 H(a) \sqrt{1 + \left[\frac{3\Omega_{b,0}}{4\Omega_{\gamma,0}}\right] a}},
\end{eqnarray}

where the following values have been taken: $\Omega_{b,0} = 0.02242$ \cite{Planck:2018vyg}; $T_0 = 2.7255 \, \text{K}$ \cite{Fixsen_2009temcmb}; and a fiducial value of $r_{s, \text{fid}}(z_d) = 147.78 \, \text{Mpc}$. The related estimate of $\chi^{2}_{BAO}$ is given by \cite{Conley_2010Apjs}
\begin{equation}\label{chisquare_bao}
\chi^2_{\text{BAO}}(\Theta) = \left(\Delta \mathcal{F}(z_i, \Theta)\right)^T C^{-1}_{\text{BAO}} \Delta \mathcal{F}(z_i, \Theta)\,.
\end{equation}

The covariance matrix for all considered BAO observations is denoted as $C_{\text{BAO}}$, and $\Delta \mathcal{F}(z_i, \Theta)$ is defined as $\mathcal{F}(z_i, \Theta) - \mathcal{F}_{\text{obs}}(z_i)$.\\


\section{Cosmological models} \label{cosmologicalmodels}

In this section, we present and examine the results based on the approach described in Sec.~\ref{SEC_observational_data} and utilizing the observational data mentioned earlier. Each subsection emphasizes the most promising models of potential functions $V(\phi)$, featuring contour plots of the constrained parameters with uncertainties $1\sigma$ and $2\sigma$ and tables displaying the results. These models have become significant in literature and are often examined due to their ability to reflect the cosmological history closely. In all the tables and posterior plots, we present results for the Hubble constant $H_0$, the current matter density parameter $\Omega_{m0}$, and the model parameters. This setup will enable us to evaluate how various independent data sets and cosmological models influence the Hubble tension.
\subsection{Model-I: Power Law Potential}
We have consider the power law potential \cite{Copeland:2006wr}, where the potential function is defined as 
\begin{equation}\label{model_I}
 V(\phi)=V_{0} \phi^{n}\,.   
\end{equation}

This specific form of the potential is defined by its reliance on the scalar field $\phi$ raised to the power of $n$, with $V_0$ serving as the constant coefficient. This potential is frequently used in inflation and dark energy models, as it accommodates various dynamical behaviors depending on the choice of $n$. Many authors have studied the cosmological evolution history by considering a power-law potential function \cite{Hossain:2024prd, Duchaniya_2024CQG, Kadam_2024Aop}. In our investigation, we will examine the consequences of this potential on cosmological development and the associated observational data set. For this potential, the Hubble parameter $H(z)$ is expressed as a function of redshift $z$ 
\begin{equation}\label{modelI_Hz}
H^2(z)=\frac{3H_0^2(\Omega_{m0} (1+z)^3+\Omega_{r0} (1+z)^4)+8\pi G V_0 \phi^n}{3-4 \pi G (1+z)^2 \left(\frac{d\phi}{dz}\right)^2}\,,    
\end{equation}
By utilizing this potential, we can determine the value of \( V_0 \)
\begin{eqnarray}\label{modelI_LCDMlimit}
V_0=\frac{3 H_0^2 (1-\Omega_{m0}-\Omega_{r0})}{8\pi G}\,.  
\end{eqnarray}

The Hubble parameter function described in Eq.~ \eqref{modelI_Hz} reduced to the standard $\Lambda$CDM model when we set the model parameter $n=0$ and utilize the corresponding $V_{0}$ value as indicated in Eq.~ \eqref{modelI_LCDMlimit}. In this analysis, we have fixed the parameter $V_0$ due to difficulties in adequately constraining it across the available data sets. Subsequently, we employed the MCMC approach to constrain the model parameters, specifically $H_0$, $\Omega_{m0}$, and $n$. In this scenario, the Hubble parameter $ H(z)$ \eqref{modelI_Hz} is influenced by the scalar field function $\phi(z)$ and its derivative $ \phi'(z) $. Consequently, we have derived the solutions for  $\phi(z) $ and $ \phi'(z)$ from the Klein-Gordon Eq.~\eqref{redshift_Klein_gordan_equation}. To find the solution to the second-order non-linear differential equation, we utilized numerical methods to determine $\phi(z) $ and $ \phi'(z)$ based on Eq.~ \eqref{redshift_Klein_gordan_equation}. We simultaneously solved the Hubble parameter $H(z)$ and the Klein-Gordon equation for each redshift $z_i$ in the range $0<z<2.4$ to obtain the posterior distribution and the best-fit values of the model parameters for various combinations of data sets using the MCMC technique. 

\begin{figure}[htbp]
 \centering
 \includegraphics[width=80mm]{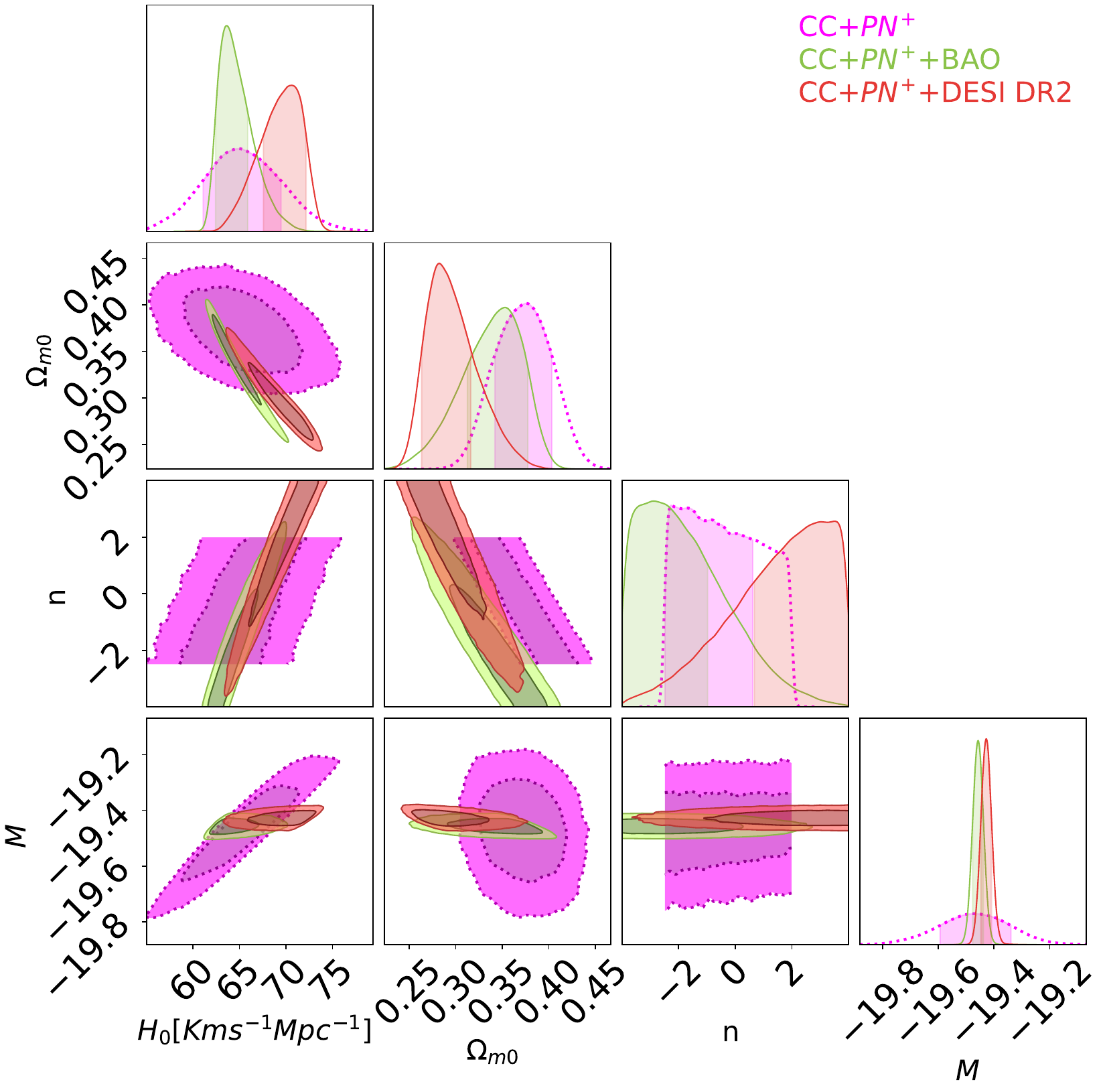}
  \includegraphics[width=80mm]{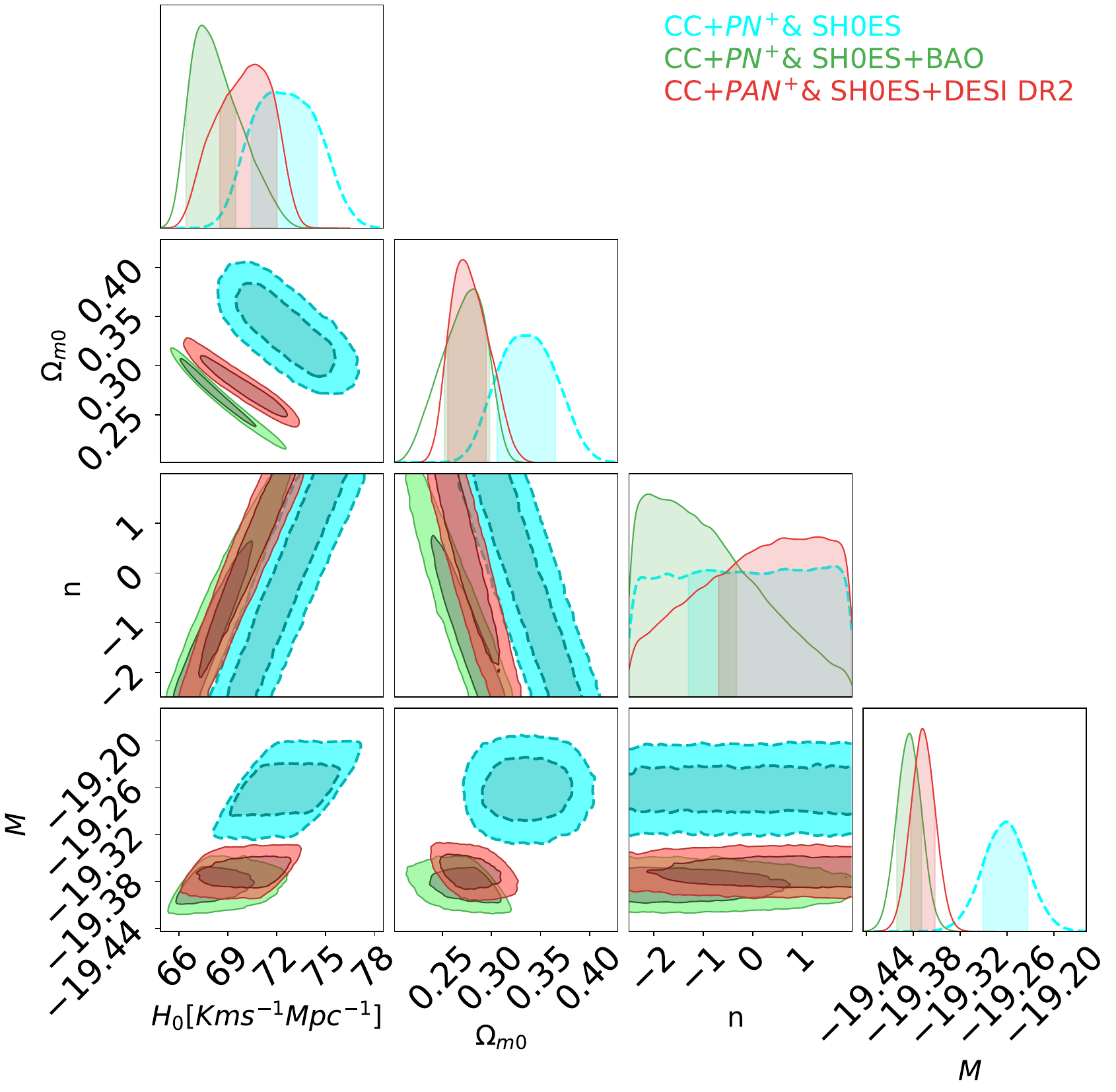}
 \caption{Left: The confidence intervals and posterior distributions for the power law model obtained from the combined data sets CC, PN$^{+}$, BAO, and DESI DR2. Right: The confidence intervals and posterior distributions for the power law model utilizing the data sets CC, PN$^{+}$, BAO, and DESI DR2.} \label{powerlawCCSNBAO}
 \end{figure}

In Fig.~\eqref{powerlawCCSNBAO}, we have defined the confidence regions along with the posteriors distribution for different combinations of observational data sets. Particularly, the figure displays the outcomes for data collections that consist of either the PN$^+$ catalog or the PN$^{+}$ \& SH0ES. The inner contours show the 68\% confidence level, whereas the outer contours depict the 95\% confidence level. This visual illustration aids in a comprehensive assessment of the uncertainties and correlations of the parameters. Conversely, the contour plots for the CC+PN$^{+}$ and the CC+PN$^{+}$ \& SH0ES data set combinations reveal a degeneracy between the $H_0$, $\Omega_{m0}$ and the $n$ parameter. 
Nonetheless, when the BAO and DESI DR2 data sets are incorporated, this degeneracy is broken, exposing an anti-correlation between parameters. However, the intensity of this anti-correlation is less prominent in the data sets that incorporate the BAO and DESI DR2. On the other hand, the  precise values for the model parameters include Nuisance parameter listed in Table~\ref{powerlaw_outputs}. It becomes evident that the $H_0$ values for data set combinations that encompass CC+PN$^{+}$ \& SH0ES are comparatively greater than their associated $H_0$ values. This observation aligns with the elevated $H_0$ value reported by the SH0ES team (R22), which states $H_0 = 73.30 \pm 1.04 \,\text{km s}^{-1} \, \text{Mpc}^{-1}$ \cite{Riess_2022panplus}. The findings indicate that the greatest values of $H_0$ are achieved for the CC+PN$^{+}$ \& SH0ES, yielding a value of $H_0 = 71.7^{+2.8}_{-1.3} \,\text{km s}^{-1} \, \text{Mpc}^{-1}$, due to the SH0ES data points. This adjustment shifts \(H_0\) to the range of 71 to 74. On the other hand, the minimum value for the H0 parameter is obtained from the CC+PN$^{+}$ data sets, mainly impacted by the BAO data that pertains to the conditions of the early Universe.The upcoming section will present a more in-depth statistical examination of these results and a comparison to the $\Lambda$CDM model.

\begin{table}
 \renewcommand{\arraystretch}{2.2} 
    \centering
    \caption{The table displays outcomes for the power law model, with the initial column enumerating the combinations of data sets. The second column indicates the restrictions on the Hubble constant $H_0$, while the third and fourth columns show the values for the matter density $\Omega_{m0}$ and the model parameter $n$, respectively. The final column illustrates the Nuisance parameter $M$.}
    \label{powerlaw_outputs}
    \begin{tabular}{ccccc}
        \hline
		Data sets & $H_0\,\text{km} \, \text{s}^{-1} \text{Mpc}^{-1}$ & $\Omega_{m0}$ & $n$ & $M$ \\ 
		\hline
		CC+$PN^{+}$ & $65.0^{+4.5}_{-3.9}$ & $0.376^{+0.027}_{-0.034}$ & $-2.36^{+2.98}_{-0.13}$ & $-19.47\pm 0.13$ \\
		CC+$PN^{+}$ +BAO & $63.6^{+2.3}_{-1.1}$ & $0.353^{+0.024}_{-0.041}$ & $-2.9^{+1.9}_{-1.1}$ & $-19.457\pm 0.019$\\ 
		CC+$PN^{+}$+DESI DR2 & $70.5^{+1.7}_{-2.9}$ & $0.282^{+0.034}_{-0.019}$ & $3.52^{+0.46}_{-2.84}$ & $-19.428\pm 0.019$\\ 
          \cline{1-5}
		CC+$PN^{+}\&$ SH0ES & $71.7^{+2.8}_{-1.3}$ & $0.337^{+0.028}_{-0.031}$ & $1.64^{+0.34}_{-2.94}$ & $-19.260^{+0.027}_{-0.030}$  \\ 
		CC+$PN^{+}\&$ SH0ES+BAO & $67.39^{+2.08}_{-0.96}$ & $0.283^{+0.015}_{-0.030}$ & $-2.12^{+1.79}_{-0.37}$ & $-19.384^{+0.015}_{-0.017}$ \\ 
		CC+$PN^{+}\&$ SH0ES+DESI DR2 & $70.7^{+1.3}_{-2.2}$ & $0.271^{+0.024}_{-0.015}$ & $1.28^{+0.71}_{-1.97}$ & $-19.368^{+0.016}_{-0.015}$\\ 
		\hline
    \end{tabular}
\end{table}

\begin{table}[htbp]
\renewcommand{\arraystretch}{2}
    \centering
    \caption{This table presents a statistical comparison between the selected model and the standard $\Lambda$CDM model. Further information about the $\Lambda$CDM model can be found in {\color{blue}Appendix}. The first column lists the data sets, including the $H_0$ priors. The second column shows the values of $\chi^{2}_{\text{min}}$. The third and fourth columns columns demonstrate the values for $\Delta \text{AIC}$ and $\Delta \text{BIC}$.}
    \label{powerlaw_outputAICBIC}
      \begin{tabular}{cccc}
        \hline
		Data set& $\chi^{2}_{min}$ &$\Delta$AIC &$\Delta$BIC \\ 
		\hline
		CC+$PN^{+}$&1792.2 &1.9 &3.2 \\ 		
        CC+$PN^{+}$+BAO& 1831.29& 15.54&16.75\\ 
		CC+$PN^{+}$+DESI DR2&1817.95 & 4.91&6.13\\ 
		\cline{1-4}
       CC+$PN^{+}\& SH0ES$&1539.28 & 2.06&3.3 \\ 
		CC+$PN^{+}\& SH0ES$+BAO& 1593.85 & 28.68&29.92\\ 
		CC+$PN^{+}\& SH0ES$+DESI DR2&1577.07 &3.57  &4.81\\
  \hline
    \end{tabular}
\end{table}
\subsection{Model-II: Quadratic Potential}
We have consider the quadratic model \cite{Dutta_2009quadratic}, which serves as a specific instance of the power law model, where the potential function is characterized as
\begin{equation}\label{model_I_qua}
 V(\phi)=V_{0} \phi^{2}\,.   
\end{equation}

We consider a specific realization of the power-law scalar field potential by choosing the quadratic form, $V(\phi)=V_0\phi^2$, which corresponds to the case $(n=2)$ in the general power-law potential $V(\phi)=V_0\phi^n$. The quadratic potential represents one of the simplest and most widely studied scalar field models in cosmology due to its minimal functional form and well-defined dynamical behavior. This choice allows us to investigate the impact of a canonical scalar field with a mass-like potential on the late-time expansion history of the Universe. Although the general power-law potential provides a broader parameter space, the quadratic case serves as an important benchmark model for understanding how scalar field dynamics address late-time cosmological phenomena, including deviations from the standard $\Lambda$CDM scenario and possible tensions in cosmological observations.  For this potential, the Hubble parameter $H(z)$ is expressed as a function of redshift $z$. 
\begin{equation}\label{modelquadratic_Hz}
H^2(z)=\frac{3H_0^2(\Omega_{m0} (1+z)^3+\Omega_{r0} (1+z)^4)+8\pi G V_0 \phi^2}{3-4 \pi G (1+z)^2 \left(\frac{d\phi}{dz}\right)^2}\,,    
\end{equation}

\begin{figure}[htbp]
 \centering
 \includegraphics[width=85mm]{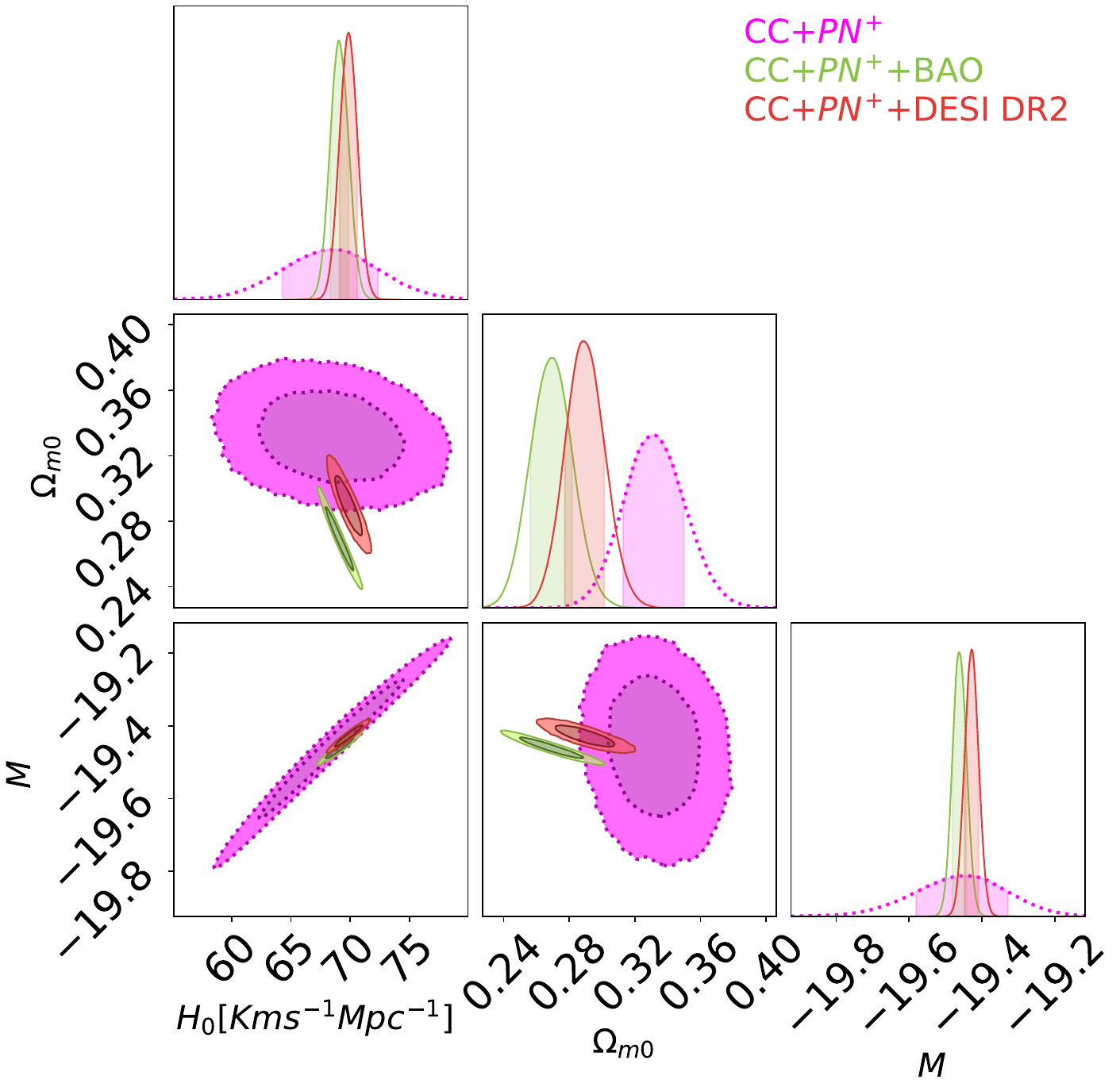}
  \includegraphics[width=85mm]{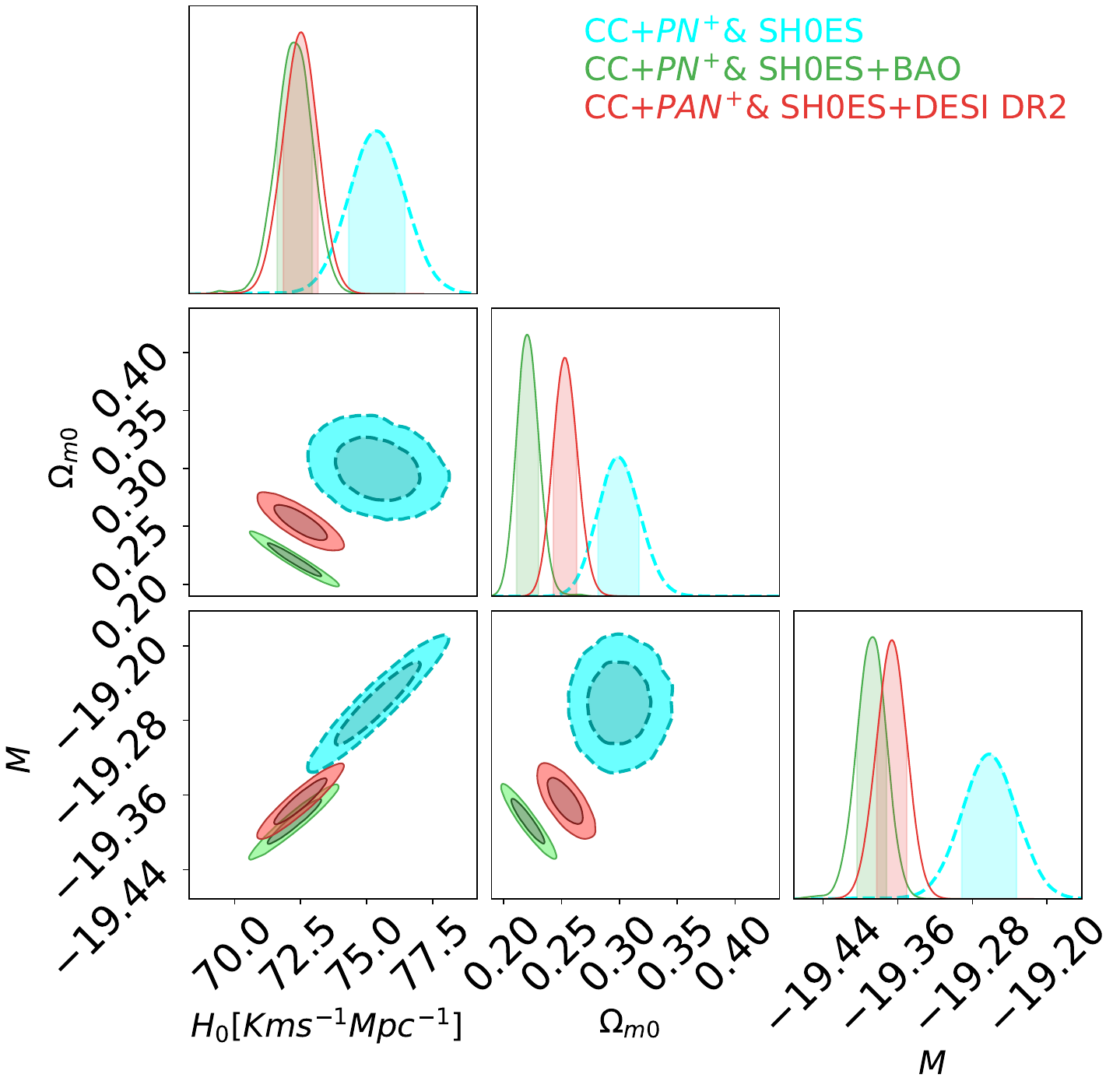}
 \caption{Left: The confidence intervals and posterior distributions for the quadratic model obtained from the combined data sets CC, PN$^{+}$, BAO, and DESI DR2. Right: The confidence intervals and posterior distributions for the quadratic model utilizing the data sets CC, PN$^{+}$, BAO, and DESI DR2.} \label{power2lawCCSNBAO}
 \end{figure}

Fig.~\eqref{power2lawCCSNBAO} displays the marginalized one-dimensional posterior distributions along with the two-dimensional confidence contours (68\% and 95\% confidence levels) for the cosmological parameters ($H_0$, $\Omega_{mo}$, M) derived from various combinations of observational data sets. On the other hand, the exact values for the model parameters, including the Nuisance parameter, are shown in Table~\ref{power2law_outputs}. Similar to Model-I, we obtain the highest value of $H_0$ for the combined data set CC+PN$^{+}$\& SH0ES, while the lowest value is obtained for the combination CC+PN$^{+}$. However, in the present quadratic potential model, the inferred values of $H_0$ are slightly higher compared to Model-I. This behavior can be attributed to the specific dynamics introduced by the quadratic potential $V(\phi)=V_0\phi^2$, which modifies the scalar field evolution and its contribution to the late-time expansion history of the Universe. The quadratic form yields a different equation-of-state evolution than the general power-law case, allowing the scalar field to exhibit a distinct dark energy behavior that can partially alter the expansion rate inferred from observational data. Consequently, the model shows a mild preference for higher $H_0$ values while remaining consistent with cosmological observations. The findings indicate that the greatest values of $H_0$ are achieved for the CC+PN$^{+}$ \& SH0ES, yielding a value of  $H_0 = 75.4\pm 1.1 \,\text{km s}^{-1} \, \text{Mpc}^{-1}$ and lowest value $H_0 = 68.5^{+3.9}_{-4.2} \,\text{km s}^{-1} \, \text{Mpc}^{-1}$  for CC+PN$^{+}$. In this context, it is important to note that for CC+PN$^{+}$ and SH0ES, the $\Omega_{m0}$ parameter reaches a lower minimum value in comparison to CC+PN$^{+}$. This indicates that a significant portion of the universe's energy is expressed as an effective dark energy, which aligns with the increased value of $H_0$.
\begin{table}
 \renewcommand{\arraystretch}{2.2} 
    \centering
    \caption{The table displays outcomes for the quadratic model, with the initial column enumerating the combinations of data sets. The second column indicates the restrictions on the Hubble constant $H_0$, while the third and fourth columns show the values for the matter density $\Omega_{m0}$ and the Nuisance parameter $M$.}
    \label{power2law_outputs}
    \begin{tabular}{cccc}
        \hline
		Data sets & $H_0\,\text{km} \, \text{s}^{-1} \text{Mpc}^{-1}$ & $\Omega_{m0}$ & $M$ \\ 
		\hline
		CC+$PN^{+}$ & $68.5^{+3.9}_{-4.2}$ & $0.332^{+0.018}_{-0.019}$ & $-19.44^{+0.12}_{-0.14}$\\
		CC+$PN^{+}$ +BAO & $69.05^{+0.82}_{-0.73}$ & $0.270^{+0.012}_{-0.013}$ & $-19.462^{+0.020}_{-0.018}$ \\ 
		CC+$PN^{+}$+DESI DR2 & $69.85^{+0.75}_{-0.76}$ & $0.288^{+0.013}_{-0.011}$ & $-19.427^{+0.018}_{-0.020}$ \\ 
          \cline{1-4}
		CC+$PN^{+}\&$ SH0ES & $75.4\pm 1.1$ & $0.299^{+0.018}_{-0.017}$ & $-19.262^{+0.029}_{-0.030}$ \\ 
		CC+$PN^{+}\&$ SH0ES+BAO & $72.20^{+0.75}_{-0.59}$ & $ 0.220^{+0.009}_{-0.009} $ & $-19.386^{+0.015}_{-0.017}$\\ 
		CC+$PN^{+}\&$ SH0ES+DESI DR2 & $72.51^{+0.65}_{-0.67}$ & $ 0.252^{+0.010}_{-0.009}$ & $-19.367^{+0.017}_{-0.016}$\\ 
		\hline
    \end{tabular}
\end{table}
\begin{table}[htbp]
\renewcommand{\arraystretch}{2}
    \centering
    \caption{This table presents a statistical comparison between the selected model and the standard $\Lambda$CDM model. Further information about the $\Lambda$CDM model can be found in {\color{blue}Appendix}. The first column lists the data sets, including the $H_0$ priors. The second column shows the values of $\chi^{2}_{\text{min}}$. The third and fourth columns columns demonstrate the values for $\Delta \text{AIC}$ and $\Delta \text{BIC}$.}
    \label{power2law_outputAICBIC}
      \begin{tabular}{cccc}
        \hline
		Data sets& $\chi^{2}_{min}$ &$\Delta$AIC &$\Delta$BIC \\ 
		\hline
		CC+$PN^{+}$&1792.94 &0.64 &0.74 \\ 
		CC+$PN^{+}$+BAO& 1836.05 &18.3 &18.3\\ 
		CC+$PN^{+}$+DESI DR2&1818.19 &3.15 &3.16\\ 
		\cline{1-4}
       CC+$PN^{+}\& SH0ES$& 1539.27 &0.05 &0.05 \\ 
		CC+$PN^{+}\& SH0ES$+BAO& 1598.22 &31.05 &31.05\\ 
		CC+$PN^{+}\& SH0ES$+DESI DR2& 1577.14 &1.64 &1.65\\
  \hline
    \end{tabular}
\end{table} 
\subsection{Model-III: Fifth power Potential}
We have considered the fifth power potential \cite{Scherrer_2008}, which again represents a particular example of the power law model, where the potential function is defined as
\begin{equation}\label{model_fifthpower}
 V(\phi)=V_{0} \phi^{5}\,.   
\end{equation}

In this section, we investigate another specific realization of the general power-law potential by considering the fifth-power potential, $V(\phi)=V_0\phi^5$, corresponding to the case ($n=5$) in the general form $V(\phi)=V_0\phi^n$. Unlike the quadratic potential, the fifth-power potential is considerably steeper for larger field values, leading to a different evolution of the scalar field and its contribution to the cosmic expansion history. The increased steepness causes the scalar field to evolve more rapidly, thereby modifying the effective dark energy dynamics during the late-time evolution of the Universe. Consequently, although the overall cosmological behavior remains qualitatively similar to that of the general power-law model, the quantitative constraints on the model parameters, particularly the Hubble constant $H_0$, differ noticeably. Comparing the three cases, the general power-law model provides the greatest flexibility through the free exponent  $n$.
In contrast, the quadratic and fifth-power potentials represent two distinct fixed realizations with different scalar-field dynamics. The quadratic potential typically yields a smoother evolution of the scalar field. In contrast, the fifth-power potential, owing to its steeper slope, can lead to stronger deviations in the expansion history and, consequently, different observational constraints. Therefore, studying these special cases allows us to assess how the choice of the potential shape influences the cosmological evolution and the inferred model parameters. For this potential, the Hubble parameter $H(z)$ is expressed as a function of redshift $z$ 

\begin{equation}\label{modelfifth_Hz}
H^2(z)=\frac{3H_0^2(\Omega_{m0} (1+z)^3+\Omega_{r0} (1+z)^4)+8\pi G V_0 \phi^5}{3-4 \pi G (1+z)^2 \left(\frac{d\phi}{dz}\right)^2}\,,    
\end{equation}

\begin{figure}[htbp]
 \centering
 \includegraphics[width=80mm]{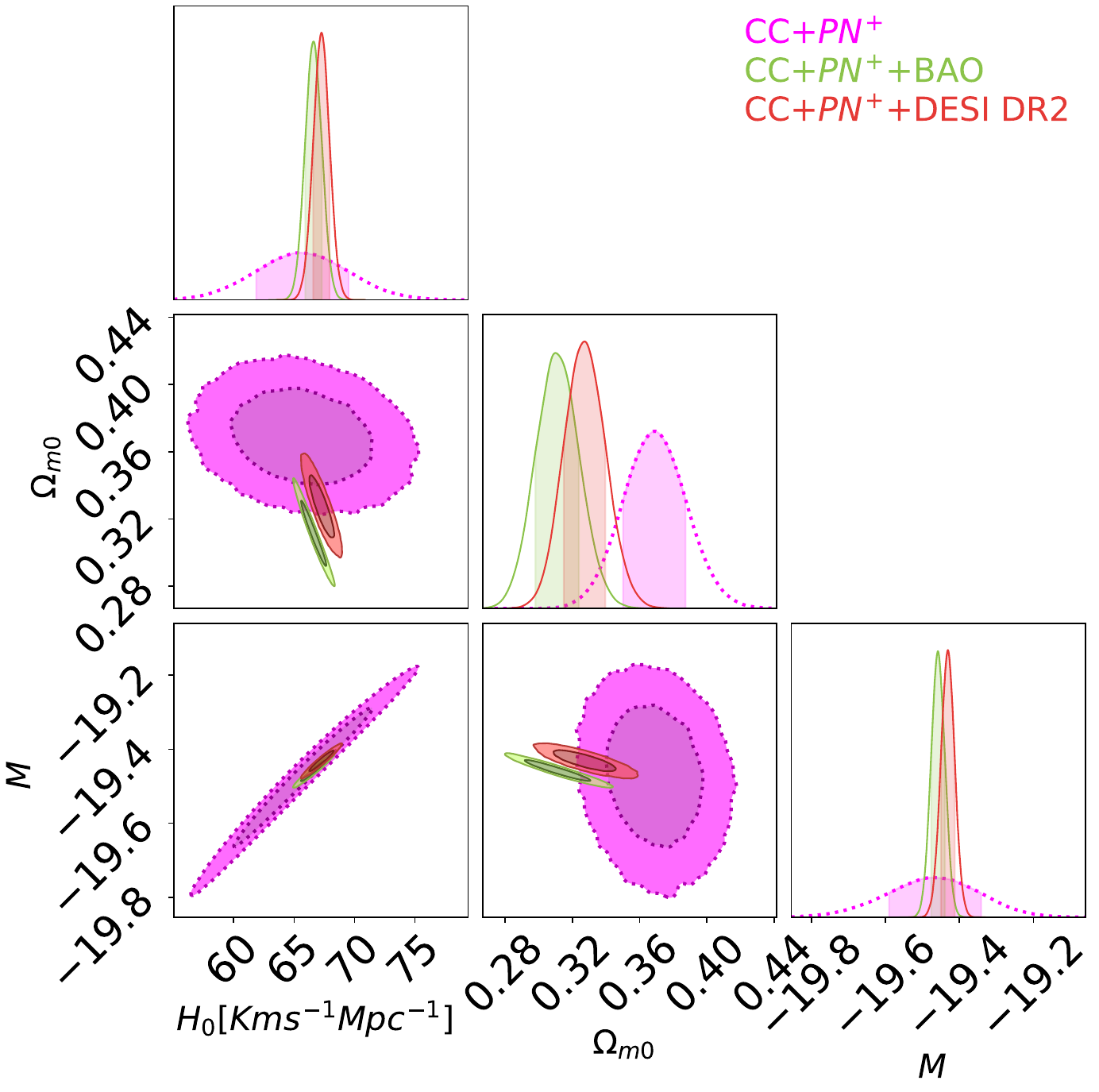}
  \includegraphics[width=80mm]{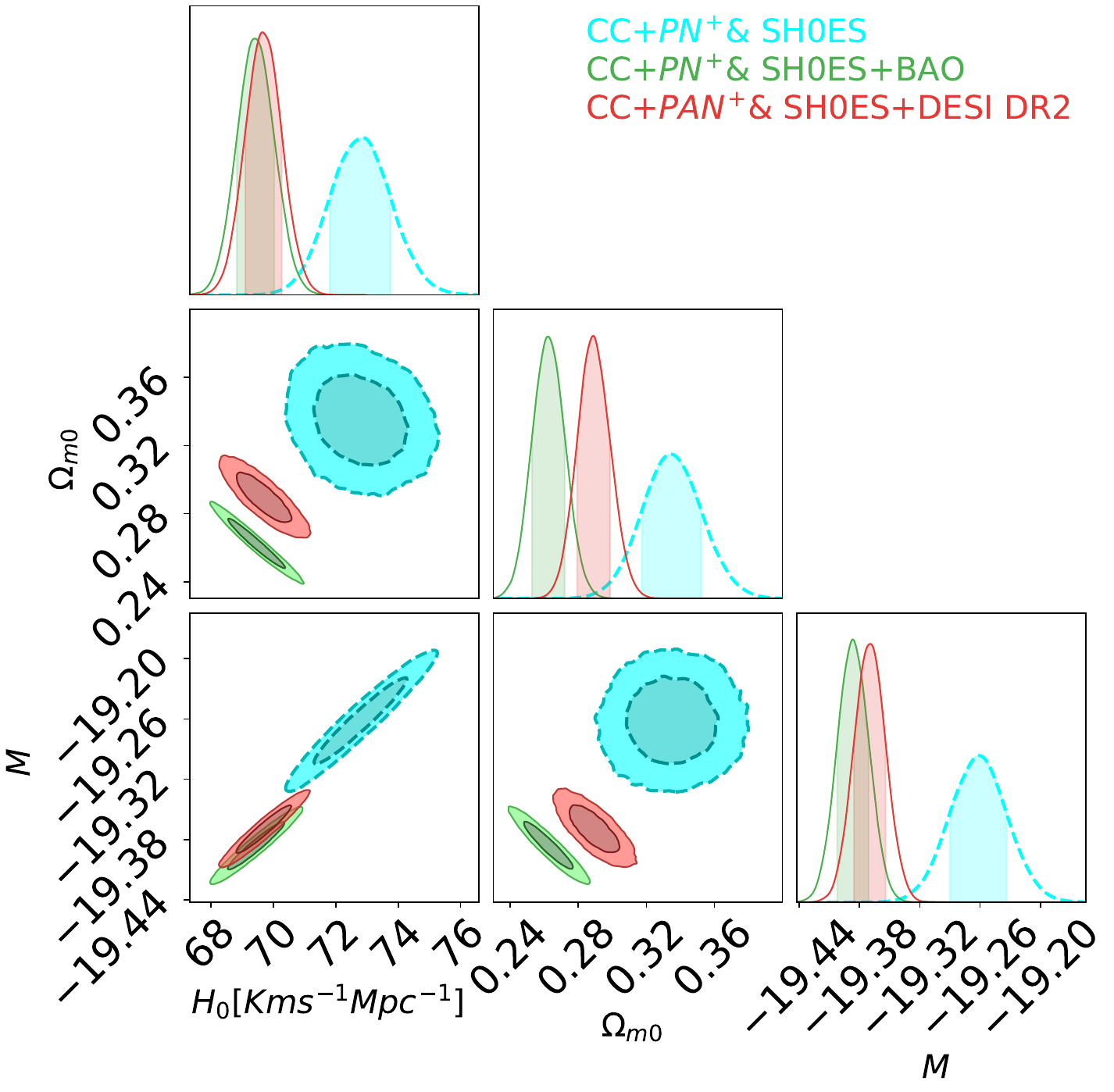}
 \caption{Left: The confidence intervals and posterior distributions for the fifth power potential model obtained from the combined data sets CC, PN$^{+}$, BAO, and DESI DR2. Right: The confidence intervals and posterior distributions for the fifth power potential model utilizing the data sets CC, PN$^{+}$, BAO, and DESI DR2.} \label{power5lawCCSNBAO}
 \end{figure}

The marginalized one-dimensional posterior distributions and the associated two-dimensional confidence contours at the 68\% and 95\% confidence levels for the cosmological parameters ($H_0$, $\Omega_{m0}, M$), obtained using different combinations of observational data sets, are shown in Fig.~\eqref{power5lawCCSNBAO}. The corresponding parameter estimates, including the nuisance parameter, are presented in Table~\ref{power5law_outputs}. Like Model-I, we achieve the maximum value of \(H_0\) for the combined dataset CC+PN$^{+}$ \& SH0ES, whereas the minimum value is derived from the combination CC+PN$^{+}$. For CC+PN$^{+}$ \& SH0ES data set combination,  we have obtained $H_0 = 72.91^{+0.85}_{-1.10} \,\text{km s}^{-1} \, \text{Mpc}^{-1}$. This observation aligns with the elevated $H_0$ value reported by the SH0ES team (R22), which states $H_0 = 73.30 \pm 1.04 \,\text{km s}^{-1} \, \text{Mpc}^{-1}$ \cite{Riess_2022panplus}. 

\begin{table}
 \renewcommand{\arraystretch}{2.2} 
    \centering
    \caption{The table displays outcomes for the fifth power potential model, with the initial column enumerating the combinations of data sets. The second column indicates the restrictions on the Hubble constant $H_0$, while the third and fourth columns show the values for the matter density $\Omega_{m0}$ and the Nuisance parameter $M$.}
    \label{power5law_outputs}
    \begin{tabular}{cccc}
        \hline
		Data sets & $H_0\,\text{km} \, \text{s}^{-1} \text{Mpc}^{-1}$ & $\Omega_{m0}$ & $M$ \\ 
		\hline
		CC+$PN^{+}$ & $65.5^{+4.1}_{-3.6}$ & $0.369^{+0.018}_{-0.019}$ & $-19.48^{+0.13}_{-0.11}$  \\
		CC+$PN^{+}$ +BAO & $66.64^{+0.66}_{-0.72}$ & $0.309^{+0.015}_{-0.011}$ & $-19.458^{+0.018}_{-0.019}$ \\ 
		CC+$PN^{+}$+DESI DR2 & $67.25^{+0.69}_{-0.67}$ & $0.327^{+0.012}_{-0.013}$ & $-19.431^{+0.018}_{-0.019}$ \\ 
          \cline{1-4}
		CC+$PN^{+}\&$ SH0ES & $72.91^{+0.85}_{-1.10}$ & $0.335^{+0.017}_{-0.018}$ & $-19.261^{+0.027}_{-0.030}$  \\ 
		CC+$PN^{+}\&$ SH0ES+BAO & $69.36^{+0.67}_{-0.53}$ & $0.261^{+0.010}_{-0.009} $ & $-19.387\pm 0.016$ \\ 
		CC+$PN^{+}\&$ SH0ES+DESI DR2 & $69.63^{+0.64}_{-0.53}$ & $0.288^{+0.0098}_{-0.0096}$ & $-19.369^{+0.015}_{-0.017}$\\ 
		\hline
    \end{tabular}
\end{table}
\begin{table}[htbp]
\renewcommand{\arraystretch}{1.1}
    \centering
    \caption{This table presents a statistical comparison between the selected model and the standard $\Lambda$CDM model. Further information about the $\Lambda$CDM model can be found in {\color{blue}Appendix}. The first column lists the data sets, including the $H_0$ priors. The second column shows the values of $\chi^{2}_{\text{min}}$. The third and fourth columns columns demonstrate the values for $\Delta \text{AIC}$ and $\Delta \text{BIC}$.}
    \label{power5law_outputAICBIC}
      \begin{tabular}{cccc}
        \hline
		Data sets& $\chi^{2}_{min}$ &$\Delta$AIC &$\Delta$BIC \\ 
		\hline
		CC+$PN^{+}$& 1792.29&-0.01 & 0.09 \\ 
		CC+$PN^{+}$+BAO& 1828.21 &10.46 &10.47\\ 
		CC+$PN^{+}$+DESI DR2& 1814.79&-0.25 & -0.24 \\ 
		\cline{1-4}
       CC+$PN^{+}\& SH0ES$& 1539.21 & -0.01& 0\\ 
		CC+$PN^{+}\& SH0ES$+BAO& 1591.45 &24.28 &24.28\\ 
		CC+$PN^{+}\& SH0ES$+DESI DR2& 1575.49 & -0.01 &-0.01 \\
  \hline
    \end{tabular}
\end{table}

\subsection{Model-IV: Hyperbolic Potential function}
The hyperbolic function discussed in Ref. \cite{Urel:2000sinh} is characterized by
\begin{equation}\label{model_II}
V(\phi)=V_{0}\, sinh^{\alpha}(\phi\, \gamma) \,,    
\end{equation}

where $\alpha$, $\gamma$, and $V_0$ represent the parameters of the model. This potential is applied in the context of the early Universe, where the dynamics of the scalar field $\phi$ can trigger cosmic inflation \cite{SAHNI_2000}. According to Ref. \cite{Roy_2018pha}, this model is also suitable for examining the dark energy domain or explaining late-time acceleration of the Universe. Therefore, we have chosen a power law sinh function inspired by this. For this model, we have obtained the following form of Hubble parameter $H(z)$ 
\begin{equation}\label{modelII_Hz}
H^2(z)=\frac{3H_0^2(\Omega_{m0} (1+z)^3+\Omega_{r0} (1+z)^4)+8\pi G V_{0}\, sinh^{\alpha}(\phi\, \gamma)}{3-4 \pi G (1+z)^2 \left(\frac{d\phi}{dz}\right)^2}\,.    
\end{equation}  

\begin{figure}[htbp]
 \centering
 \includegraphics[width=75mm]{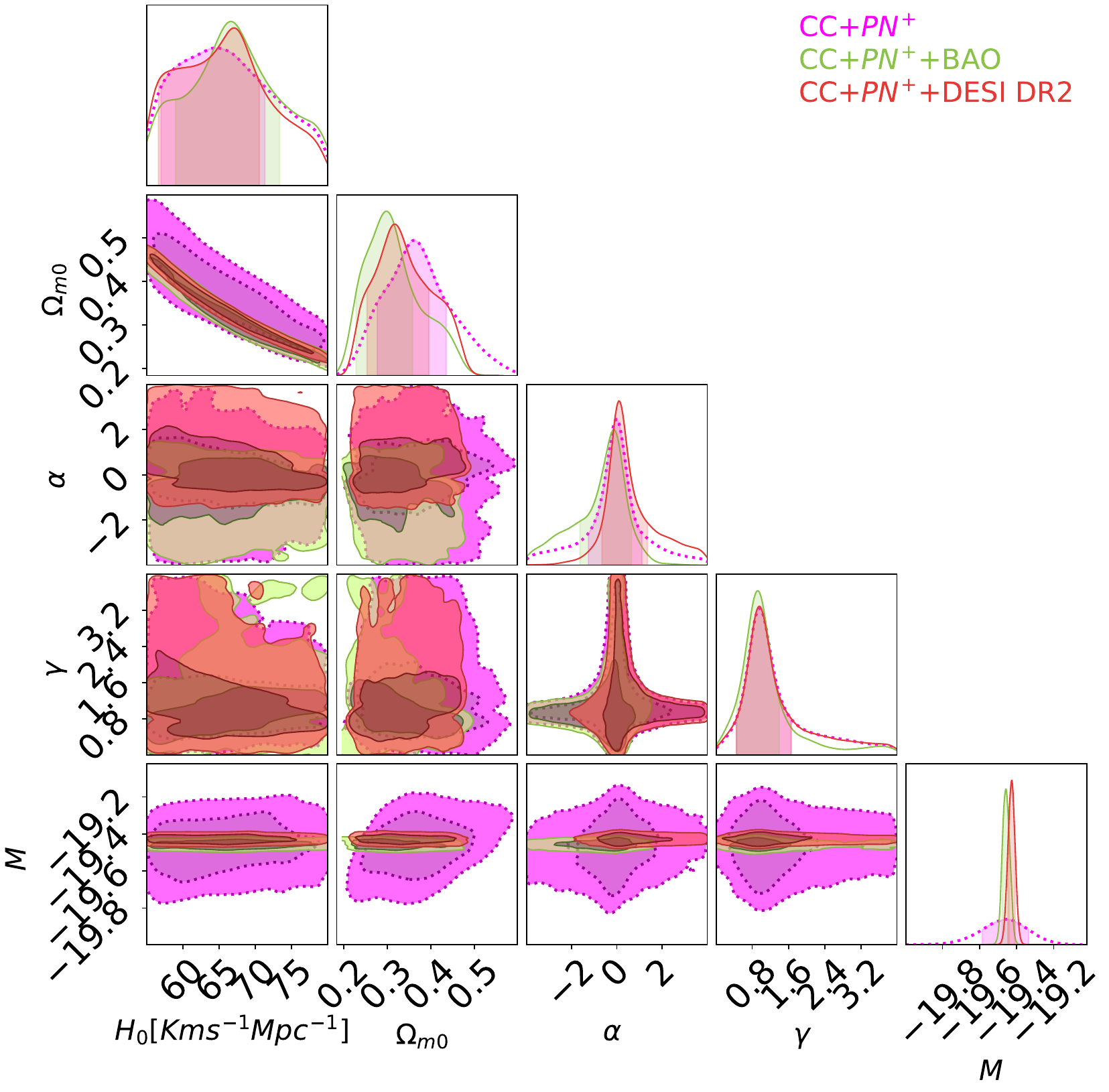}
  \includegraphics[width=75mm]{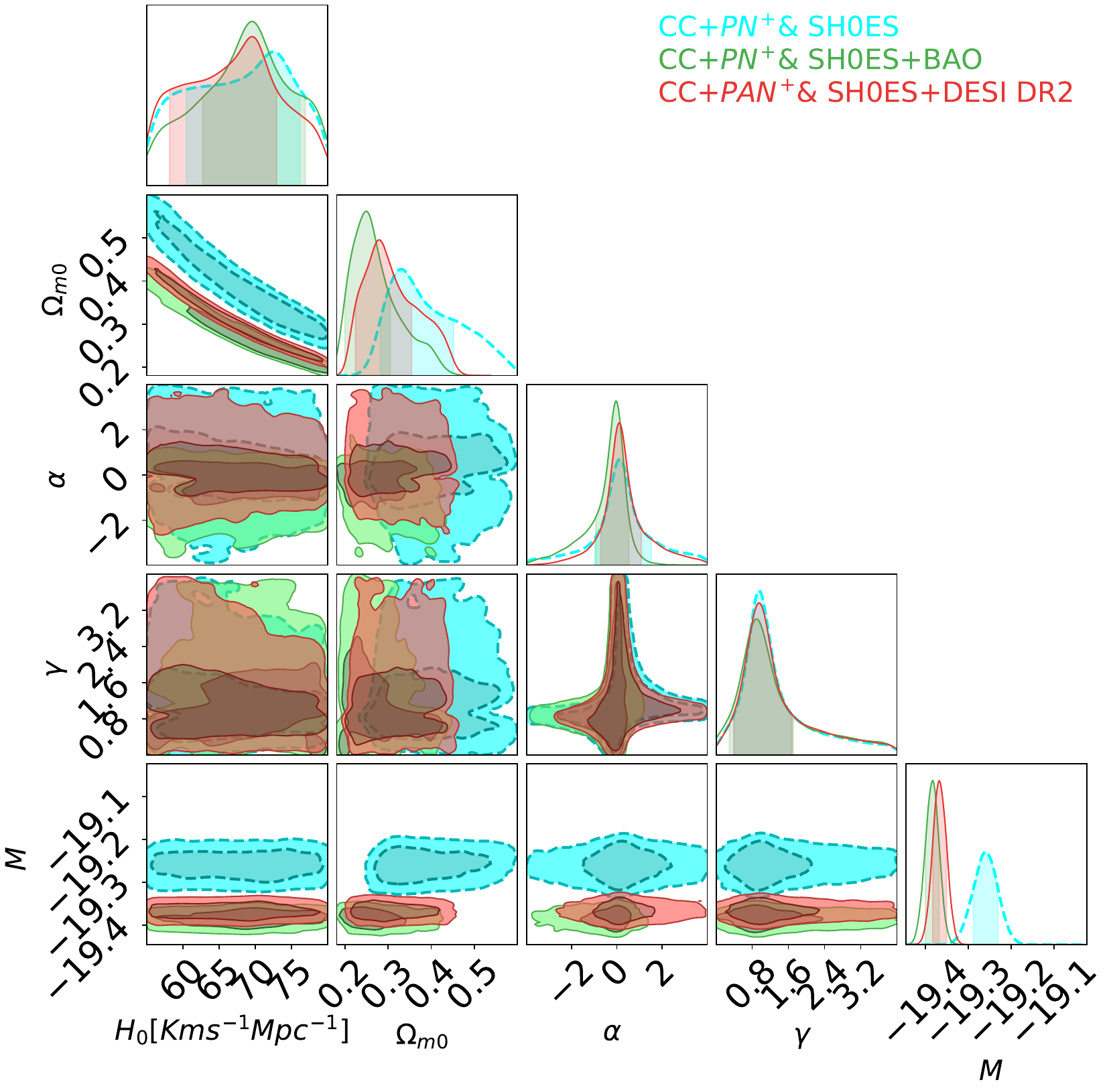}
 \caption{Left: The confidence intervals and posterior distributions for the fifth power potential model obtained from the combined data sets CC, PN$^{+}$, BAO, and DESI DR2. Right: The confidence intervals and posterior distributions for the fifth power potential model utilizing the data sets CC, PN$^{+}$, BAO, and DESI DR2.} \label{hyperbolicCCSNBAO}
 \end{figure}

This model is reduced to $\Lambda$CDM model when $\alpha = 0$, alongside the specific value of $V_0$ outlined in Eq.~\eqref{modelI_LCDMlimit}. Fig.~\eqref{hyperbolicCCSNBAO} illustrates the posterior distributions and confidence intervals of the constrained parameters for hyperbolic model. In contrast Models I-Model III yield more tightly constrained parameters, owing to the periodic-like behavior of the sinh function. As with Models I-Model III, we have also fixed the $V_0$ value, as outlined in Eq.~\eqref{modelI_LCDMlimit}. The 1$\sigma$ and 2$\sigma$ confidence contours for the hyperbolic potential model reveal distinct correlations among the model parameters. The model parameter $\alpha$ shows a mild correlation with both $H_0$ and $\Omega_{m0}$, reflecting its influence on the late-time dynamics of the Universe. The parameter $\gamma$ exhibits relatively weak correlations with the other parameters, implying that its impact is more localized and less dependent on variations in the background cosmology.

 Table~\ref{hyperbolic_outputs} displays the precise numerical values for the parameters illustrated in Fig.~\eqref{hyperbolicCCSNBAO}, encompassing the nuisance parameter M. The findings indicate that the estimated values of $H_0$ and $\Omega_{m0}$ are similar to those derived from model-I. The Hubble constant $H_0$ in this model is slightly increased, while the matter density parameter $\Omega_{m0}$ is somewhat lower than in model-I. The analysis of the data set reveals an inverse correlation between $H_0$ and $\Omega_{m0}$: when $H_0$ rises, $\Omega_{m0}$ usually falls, and vice versa, a decrease in $H_0$ is linked to an increase in $\Omega_{m0}$. For this model, the combination of the CC+$PN^{+}\&$ SH0ES data set yields a higher estimate for $H_0 $, specifically $H_0 = 72.6^{+3.6}_{-12.1}\, \text{km s}^{-1} \, \text{Mpc}^{-1} $. Additionally, the inclusion of the BAO, DESI DR2 and without SH0ES data points leads to a reduction in the estimated value of $H_0$. Additional comparisons and statistical evaluations involving the $\Lambda$CDM are presented in Sec.~\ref{modelcomparison}.
\begin{table}
\renewcommand{\arraystretch}{2}
    \centering
    \caption{The table displays results about the hyperbolic potential function, where the first column lists the combinations of data sets. The second column indicates the Hubble constant $H_0 $, while the third and fourth columns provide the values for the matter density $\Omega_{m0}$  and the model parameter $\alpha$, respectively. The fifth column represents $ \gamma$. Finally, the sixth column shows the nuisance parameter $M$.}
    \label{hyperbolic_outputs}
    \begin{tabular}{cccccc}
        \hline
		Data sets & $H_0 \,\text{km} \, \text{s}^{-1} \text{Mpc}^{-1}$ & $\Omega_{m0}$ & $\alpha$ & $\gamma$ &$M$ \\ 
		\hline
		CC+$PN^{+}$ & $64.5^{+6.8}_{-7.6}$ & $0.362^{+0.073}_{-0.086}$ & $0.0^{+1.1}_{-1.2}$ & $0.96^{+0.69}_{-0.51}$ & $-19.45^{+0.12}_{-0.13}$ \\
		CC+$PN^{+}$+BAO & $66.6^{+6.8}_{-7.6}$ & $0.298^{+0.060}_{-0.070}$ & $-0.12^{+0.77}_{-1.51}$ & $0.92\pm 0.47$ & $-19.458\pm 0.020$  \\ 
		CC+$PN^{+}$+DESI DR2 & $67.1^{+3.5}_{-10.5}$ & $0.316^{+0.080}_{-0.063}$ & $0.10^{+1.26}_{-0.75}$ & $0.96^{+0.71}_{-0.51}$ & $-19.427^{+0.018}_{-0.021}$ \\ 
          \cline{1-6}
		CC+$PN^{+}\&$ SH0ES & $72.6^{+3.6}_{-12.1}$ & $0.331^{+0.121}_{-0.049}$ & $0.1^{+1.4}_{-1.1}$ & $0.96^{+0.70}_{-0.55}$ & $-19.261^{+0.030}_{-0.028}$\\ 
		CC+$PN^{+}\&$ SH0ES+BAO &$69.5^{+7.4}_{-6.8}$ & $0.249^{+0.056}_{-0.049}$ & $-0.01^{+0.55}_{-0.92}$ & $0.91^{+0.80}_{-0.61}$ & $-19.383^{+0.015}_{-0.018}$  \\ 
		CC+$PN^{+}\&$ SH0ES+DESI DR2 & $69.5^{+3.4}_{-11.4}$ & $0.280^{+0.075}_{-0.055}$ & $0.11^{+0.99}_{-0.81}$ & $0.96^{+0.73}_{-0.57}$ & $-19.369^{+0.018}_{-0.015}$  \\ 
		\hline
    \end{tabular}
\end{table}
\begin{table}[htbp]
 \renewcommand{\arraystretch}{1.2}
    \centering
    \caption{This table provides a statistical comparison between the chosen model and the standard $\Lambda$CDM model. More details about the $\Lambda$CDM model can be found in {\color{blue}Appendix}. The first column displays the data sets, which include the $H_0$ priors. The second column presents the values of $\chi^{2}_{\text{min}}$. The third and fourth columns represent the values for $\Delta \text{AIC}$ and $\Delta \text{BIC}$.}
    \label{hyperbolic_outputAICBIC}
      \begin{tabular}{cccc}
        \hline
		Data set& $\chi^{2}_{min}$ &$\Delta$AIC &$\Delta$BIC \\ 
		\hline
		CC+$PN^{+}$&1791.87 & 3.57&6.07 \\ 		
        CC+$PN^{+}$+BAO&1831.23 &17.48 &19.89\\ 
		CC+$PN^{+}$+DESI DR2&1818.01 &6.97 &9.38\\ 
		\cline{1-4}
       CC+$PN^{+}\& SH0ES$& 1539.24&4.02 &6.5 \\ 
		CC+$PN^{+}\& SH0ES$+BAO& 1593.81 & 30.64&33.13\\ 
		CC+$PN^{+}\& SH0ES$+DESI DR2& 1577.09& 5.59 & 8.08 \\
  \hline
    \end{tabular}
\end{table}
\subsection{Model-V: Axion-like potential}
The axion-like potential function addressed in Ref. \cite{Marsh_2016axion} is formulated as follows:
\begin{equation}\label{model_III}
    V(\phi)=V_{0} \bigg(1-\cos\bigg(\frac{\phi}{\eta}\bigg)\bigg)^{\beta}\,,   
\end{equation}
where $\beta$, $\eta$, and $V_0$ represent the parameters of the model. Numerous researchers have examined the cosmological timeline, including cosmic inflation, late-time cosmology, and the Hubble constant tension issue in the context of the axion-like potential function. Poulin et al. \cite{Poulin:2018prd} analyzed how the axion-like field (ULA) influences cosmological observations as it becomes dynamic at various times in the axion-like potential function for specific values of $\beta=(1, 2, 3)$. In Ref.\cite{Poulin:2019prl}, they investigated how the early dark model can address the Hubble tension. They selected the axion-like potential function with $\beta$ values $(2,3, \infty)$. Herold and Ferreira \cite{Herold:2023prd} studied how the axion-like potential function offers a solution to the Hubble tension for the particular value of $\beta=3$. Inspired by these studies, we have focused on the axion-like potential function. However, in this investigation, we explore without fixing the value of $\beta$. We have determined the best-fit value for all these model parameters through various combinations of data sets. For the axion-like potential function, we have derived the following $H(z)$
\begin{align}\label{modelIII_Hz}
H^2(z)=\frac{3H_0^2(\Omega_{m0} (1+z)^3+\Omega_{r0} (1+z)^4)+8\pi G V_0 \bigg(1-\cos\bigg(\frac{\phi}{\eta}\bigg)\bigg)^{\beta}}{3-4 \pi G (1+z)^2 \left(\frac{d\phi}{dz}\right)^2}\,.    
\end{align}  

\begin{figure}[htbp]
 \centering
 \includegraphics[width=85mm]{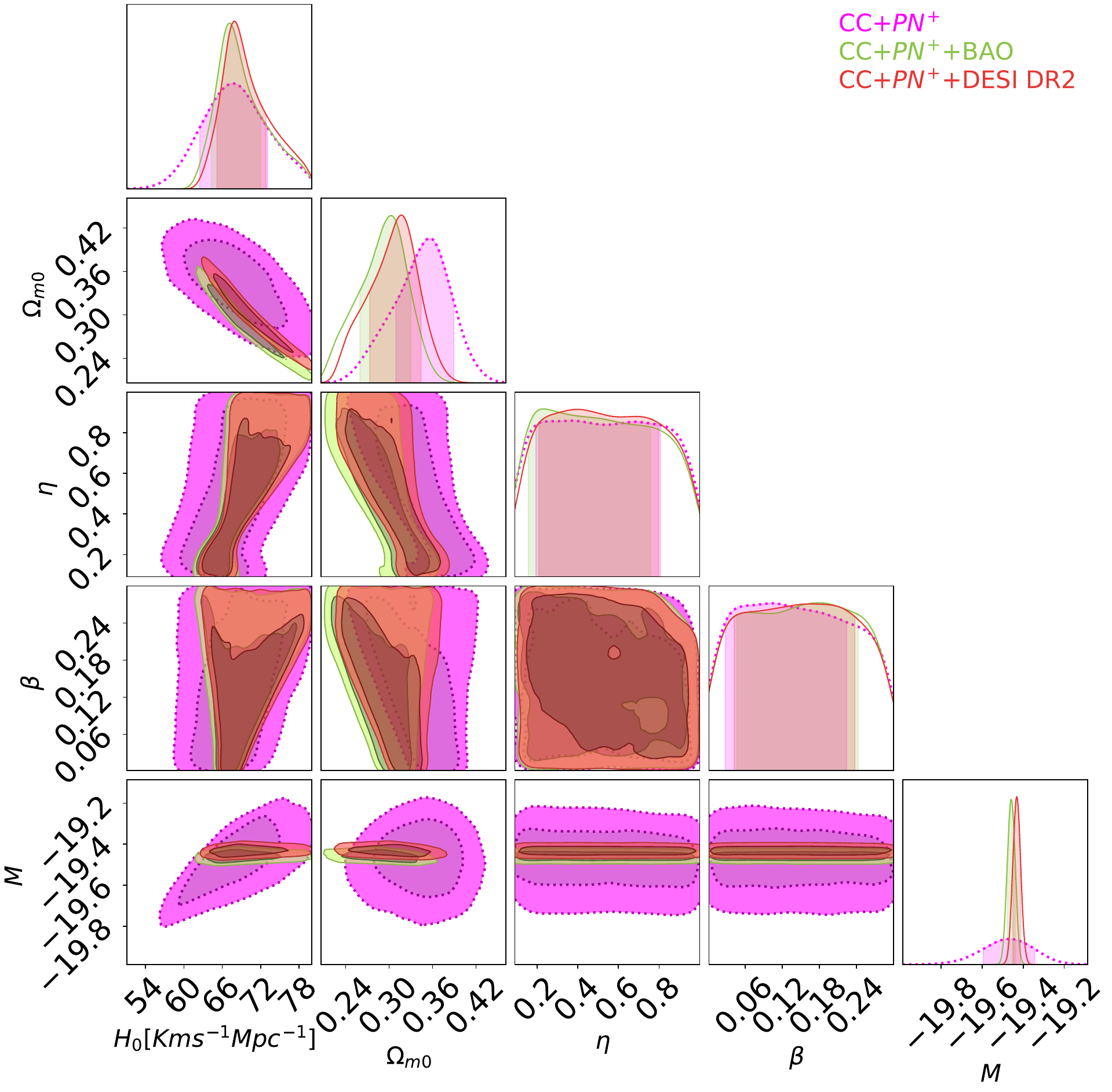}
 \includegraphics[width=85mm]{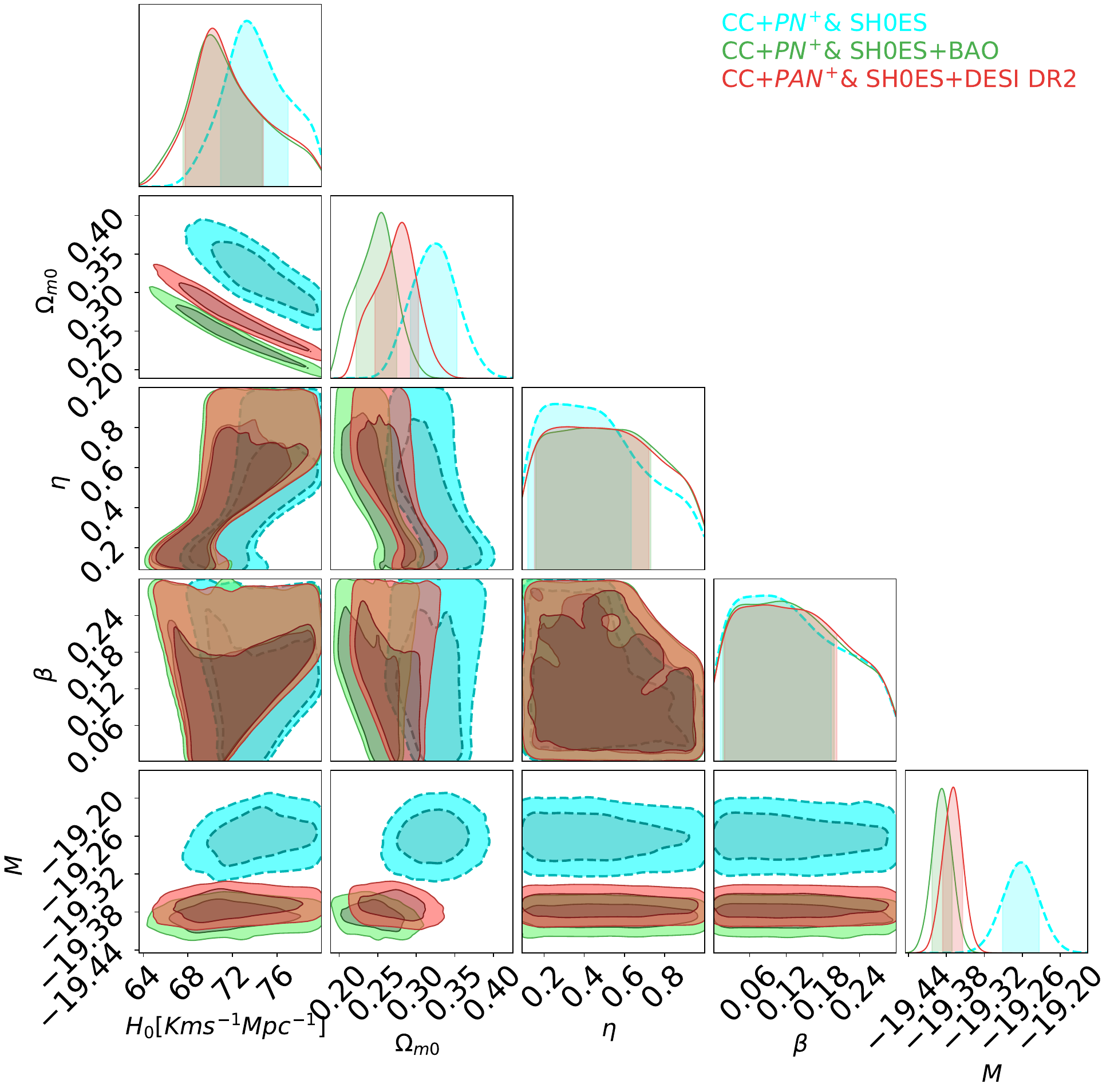}
 \caption{Left: The confidence intervals and posterior distributions for the fifth power potential model obtained from the combined data sets CC, PN$^{+}$, BAO, and DESI DR2. Right: The confidence intervals and posterior distributions for the fifth power potential model utilizing the data sets CC, PN$^{+}$, BAO, and DESI DR2.} \label{axionlikeCCSNBAO}
 \end{figure}

This model is reduced to the $\Lambda$CDM model when $\beta = 0$, along with the specific value of $V_0$ specified in Eq.~\eqref{modelI_LCDMlimit}. Fig.~ \eqref{axionlikeCCSNBAO} displays the posterior distributions and confidence intervals for the constrained parameters of model-V. The behavior shown in Fig.~\eqref{axionlikeCCSNBAO} is similar to model-IV. Consistent with Model-I -- Model-IV, we have also fixed the $V_0$ value, as described in Eq.\eqref{modelI_LCDMlimit}. The 1$\sigma$ and 2$\sigma$ confidence contours for the axion-like model highlight several important parameter correlations.  The model parameter $\beta$, associated with the dynamics of the axion-like field, shows relatively weak correlations with both $H_0$ and $\Omega_{m0}$, implying that current data places only mild constraints on its value. The  $\eta$ exhibits a mild to moderate positive correlation with $H_0$, suggesting that increased contributions from early dark energy support higher values of the Hubble constant. This trend aligns with the underlying motivation of early dark energy models in addressing the $H_0$ tension. 

Table~\ref{Axion_outputs} provides the precise numerical values for the parameters illustrated in Fig.~\eqref{axionlikeCCSNBAO}, including the nuisance parameter $M$. The analysis reveals that the model parameters $\beta$ and $\eta$ are tightly constrained, as shown by the contour plots. The $H_0$ estimate derived from the CC+PN$^{+}\&$ SH0ES combination, aligns closely with the elevated $H_0$ reported by the SH0ES team (R22), specifically $H_0 = 73.30 \pm 1.04\, \text{km s}^{-1} \, \text{Mpc}^{-1}$ \cite{Riess_2022panplus}. Consistently with Model-I -- Model-IV, the incorporation of the BAO and DESI DR2 data with the CC+PN$^{+}\&$ SH0ES combination yields a reduction in the $H_0$ values when compared to the CC+PN$^{+}\&$ SH0ES data set. The $H_0$ values derived from the combination of the CC+PN$^{+}\&$SH0ES+BAO data sets, along with the relevant priors, align with the higher estimates of $H_0$ reported in \cite{Planck:2018vyg}. 
\begin{table}
 \renewcommand{\arraystretch}{1.65}
    \centering
    \caption{The table displays results related to the axion-like model, where the first column identifies the combinations of data sets. The second column indicates the Hubble constant $H_0$, while the third and fourth columns provide the values for the matter density $\Omega_{m0}$ and the model parameter $\eta$, respectively. The fifth column represents $\beta$. The sixth column presents the Nuisance parameter $M$.}
    \label{Axion_outputs}
    \begin{tabular}{cccccc}
        \hline
		Data sets & $H_0\, \text{km} \, \text{s}^{-1} \text{Mpc}^{-1}$ & $\Omega_{m0}$ & $\eta$ & $\beta$ &$M$ \\ 
		\hline
		CC+$PN^{+}$ & $67.6^{+5.4}_{-5.2}$ & $0.356^{+0.034}_{-0.046}$ & $0.37^{+0.44}_{-0.18}$ & $0.103^{+0.121}_{-0.075}$ & $-19.47\pm 0.13$ \\
		CC+$PN^{+}$ +BAO & $67.2^{+4.8}_{-2.9}$ & $0.303^{+0.028}_{-0.043}$ & $0.229^{+0.529}_{-0.072}$ & $0.179^{+0.064}_{-0.132}$ & $-19.459^{+0.020}_{-0.019}$  \\  
		CC+$PN^{+}$ +DESI DR2 & $67.9^{+4.9}_{-2.7}$ & $0.317^{+0.027}_{-0.044}$ & $0.40^{+0.40}_{-0.19}$ & $0.164^{+0.073}_{-0.122}$ & $-19.431\pm 0.019$  \\ 
          \cline{1-6}
		CC+$PN^{+}\&$ SH0ES & $73.3^{+3.7}_{-2.4}$ & $0.325^{+0.027}_{-0.033}$ & $0.26^{+0.37}_{-0.14}$ & $0.087^{+0.106}_{-0.075}$ & $-19.262\pm 0.029$  \\ 
		CC+$PN^{+}\&$ SH0ES+BAO & $70.0^{+4.7}_{-2.4}$ & $0.255^{+0.020}_{-0.033}$ & $0.42^{+0.31}_{-0.26}$ & $0.111^{+0.088}_{-0.094}$ & $-19.387\pm 0.016$\\ 
		CC+$PN^{+}\&$ SH0ES+DESI DR2 & $70.2^{+4.5}_{-2.5}$ & $0.282^{+0.021}_{-0.035}$ & $0.32^{+0.41}_{-0.16}$ & $0.086^{+0.117}_{-0.067}$ & $-19.369^{+0.015}_{-0.017}$ \\ 
		\hline
    \end{tabular}
\end{table}
\begin{table}[htbp]
\renewcommand{\arraystretch}{1.2}
    \centering
    \caption{This table provides a statistical comparison between the chosen model and the standard $\Lambda$CDM model. More details about the $\Lambda$CDM model can be found in {\color{blue}Appendix}. The first column displays the data sets, which include the $H_0$ priors. The second column presents the values of $\chi^{2}_{\text{min}}$. The third and fourth columns represent the values for $\Delta \text{AIC}$ and $\Delta \text{BIC}$.}
    \label{Axion_outputAICBIC}
      \begin{tabular}{cccc}
        \hline
		Data set& $\chi^{2}_{min}$ &$\Delta$AIC &$\Delta$BIC \\ 
		\hline
		CC+$PN^{+}$& 1792.30&4 & 6.49\\ 		
        CC+$PN^{+}$+BAO&1828.04  &14.29 &16.7\\ 
		CC+$PN^{+}$+DESI DR2&1814.80 &9.76 &6.17\\ 
		\cline{1-4}
       CC+$PN^{+}\& SH0ES$& 1539.21& 3.99 &6.47 \\ 
		CC+$PN^{+}\& SH0ES$+BAO& 1591.14 &27.97 &30.46\\ 
		CC+$PN^{+}\& SH0ES$+DESI DR2& 1575.50 &4 &6.48\\
  \hline
    \end{tabular}
\end{table}


\section{Model Comparison} \label{modelcomparison}

We assess the effectiveness of every potential function and data set by calculating their corresponding minimum $\chi^{2}_{\text{min}}$ values derived from the maximum likelihood $L_{\text{max}}$, as $\chi^{2}_{\text{min}} = -2 \ln L_{\text{max}}$. We also evaluate the models relative to the standard $\Lambda$CDM using the Akaike Information Criterion (AIC), which considers both the fit quality (assessed by $\chi^{2}_{min}$) and the model's complexity (defined by the number of parameters, k). The AIC is expressed as 
\begin{equation}\label{AIC}
\text{AIC} = \chi^{2}_{min} + 2k,   
\end{equation}

In essence, a smaller  AIC value suggests that a model better fits the data while accounting for its complexity. The AIC penalizes models with more parameters, even if they show an improved fit to the data. Consequently, a model with a lower AIC is favored over one with a higher AIC, provided the difference in AIC is meaningful. Furthermore, we analyze the Bayesian Information Criterion (BIC), which resembles AIC but places greater emphasis on the model's complexity compared to AIC, and is defined as
\begin{equation}\label{BIC}
\text{BIC} = \chi^{2}_{min} + k \ln N,    
\end{equation}

In this context, $N$ denotes the number of samples in the observational data combination. The BIC aims to achieve the same objective as the AIC: to balance the model's accuracy on the data with its complexity. Nevertheless, the BIC imposes a greater penalty on models with more parameters than the AIC, due to its use of the logarithm of the sample size. As a result, the penalty for additional parameters becomes more pronounced as the sample size rises. In practical terms, evaluating BIC values between two models can help identify which model is more consistent with the data, with models exhibiting lower BIC values preferred as long as the difference is significant.

To evaluate the performance of different models with various combinations of data sets, we determine the differences in AIC and BIC between each model and the reference model, $\Lambda$CDM. The constrained parameters for the $\Lambda$CDM model corresponding to each combination of data sets are detailed in Table~\ref{LCDM_outputs} of the {\color{blue}Appendix}. Lower values of $\Delta$AIC and $\Delta$BIC indicate that the model using the selected data set aligns more closely with the $\Lambda$CDM model, reflecting superior performance. Tables~\ref{powerlaw_outputAICBIC}, \ref{power2law_outputAICBIC}, \ref{power5law_outputAICBIC} \ref{hyperbolic_outputAICBIC}, and \ref{Axion_outputAICBIC} present the values for different statistical measures, including $\chi^{2}_{\text{min}}$, $\Delta$AIC, and $\Delta$BIC for each model. The quantities $\Delta$AIC, and $\Delta$BIC can be defined as,
\begin{align}
\Delta\text{AIC} &= \Delta \chi^{2}_{min} + 2\,\Delta k\,,\\
\Delta\text{BIC} &= \Delta \chi^{2}_{min} + \Delta k\, \ln N\,.
\end{align}

In Tables~\ref{powerlaw_outputAICBIC}, \ref{power2law_outputAICBIC}, \ref{power5law_outputAICBIC} \ref{hyperbolic_outputAICBIC}, and \ref{Axion_outputAICBIC}, we observe that the combination of CC + $PN^{+}\&$ SH0ES, CC + $PN^{+}\&$ SH0ES+DESI DR2, CC + $PN^{+}$, and CC + $PN^{+}$ + DESI DR2 yields lower values for both $\Delta\text{AIC}$ and $\Delta\text{BIC}$. This suggests that this data set configuration aligns more closely with the standard $\Lambda$CDM model. Conversely, when incorporating the BAO data set alongside CC+$PN^{+}\&$ SH0ES and CC+$PN^{+}$, we see an increase in $\Delta\text{AIC}$ and $\Delta\text{BIC}$, which implies that this particular observational combination provides weaker support for the model compared to the $\Lambda$CDM model.

\begin{figure}[htbp]
 \centering
 \includegraphics[width=150mm]{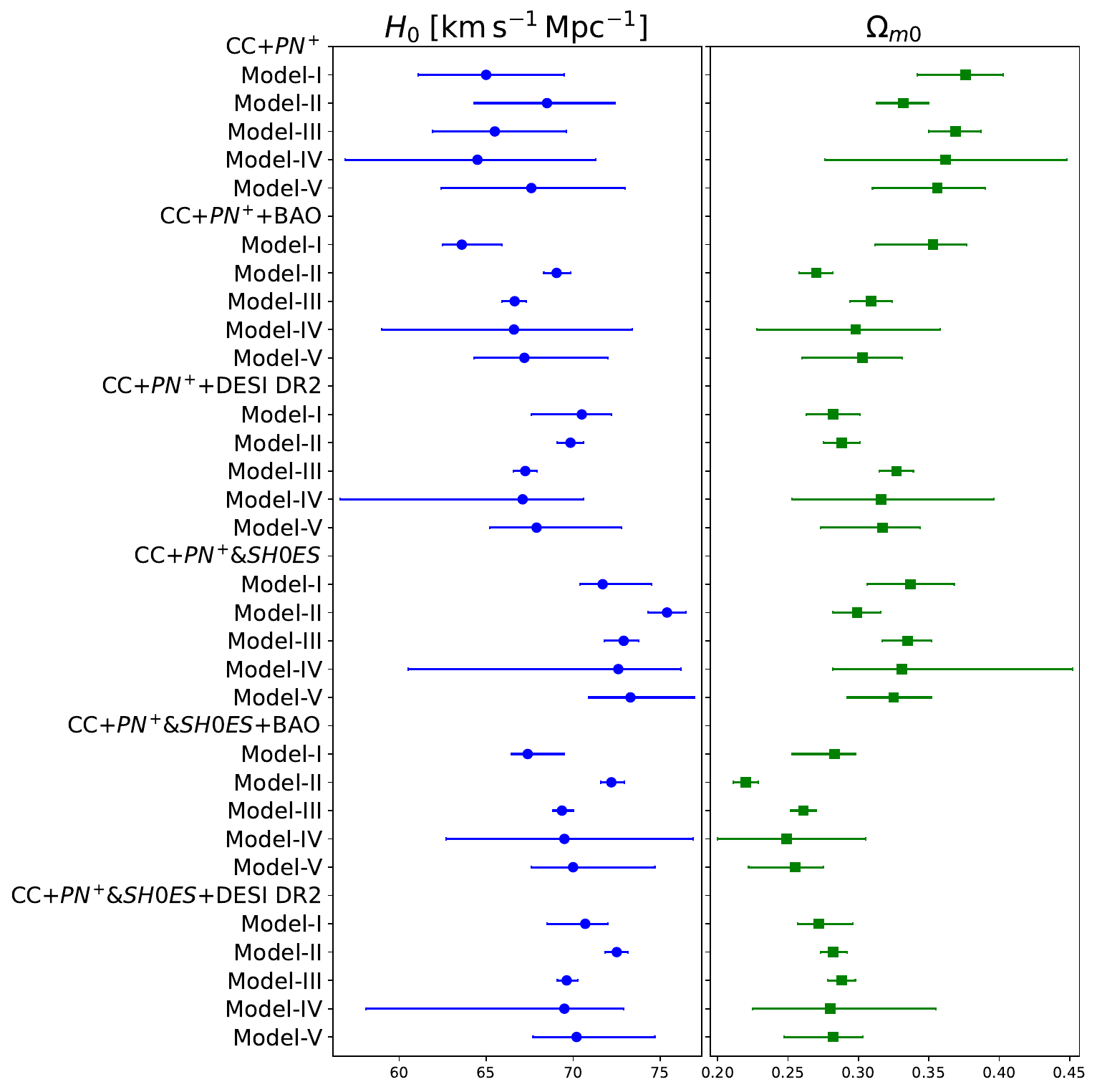} 
 \caption{The whisker plot illustrates the distribution parameters $H_0$, $ \Omega_{m0}$ for Models I-V obtained from different combinations of observational data sets. The central marker denotes the best-fit value, while the horizontal bars represent the corresponding confidence intervals.} \label{whisker}
 \end{figure}

To simplify the cross-analysis among the five models and the various observational data combinations, we present the whisker plots in Fig.~\ref{whisker}, illustrating the marginalized constraints on the cosmological parameters $H_0$ and $\Omega_{m0}$. For all five models, the highest values of $H_0$ are consistently obtained from the CC+PN$^{+}$\&SH0ES data combination. However, the inclusion of the BAO or DESI DR2 data with CC+PN$^{+}$\&SH0ES systematically shifts the inferred $H_0$ toward lower values. This behavior reflects the constraining power of BAO and DESI DR2 measurements, which probe the expansion history through early-Universe standard rulers and therefore favor comparatively lower values of the Hubble constant. Furthermore, an inverse correlation between $H_0$ and $\Omega_{m0}$ is evident across all models: data combinations that yield higher values of $H_0$ tend to prefer lower values of $\Omega_{m0}$, whereas lower values of $H_0$ are accompanied by higher estimates of $\Omega_{m0}$. This anti-correlation reflects the well-known degeneracy between the present-day expansion rate and the matter density parameter. A higher value of $H_0$ together with a lower value of $\Omega_{m0}$ indicates that the current accelerated expansion of the Universe is more strongly driven by the dark energy component, while a lower $H_0$ and a higher $\Omega_{m0}$ suggest a relatively greater contribution of matter to the cosmic expansion history.

\section{Discussions and Conclusion} \label{conclusion}

In recent years, scalar field theories have shown great promise in meeting the growing challenges in the observational predictions of $\Lambda$CDM. This has taken various forms in terms of coupled and uncoupled scalar fields, as well as both early and late time centered fields. In this work, we explore the possible observations and impacts of a late time acting scalar field by adopting five physically motivated potentials and using data sets located in the late Universe.

The models explored in this work are driven by a power-law potential and a special case of the power law function. We have also taken a quadratic model, a fifth-power potential model, a hyperbolic sinh function, and an axion-like function, which encompass many of the possible general behaviors of interest in the literature. The power-law model captures the general trend of models that have a healthy $\Lambda$CDM limit but may deviate when the scalar field takes large values, such as in the late Universe. The hyperbolic model produces a smooth transition in the sign of the potential for different values of the scalar field, rescaled by the $\gamma$ model parameter. As for the final axion-like function, this incorporates possible oscillatory behavior in this regime of the Universe. 

These potentials are constrained by the consideration of CC, PN$^{+}$\& SH0ES, PN$^{+}$, DESI DR2, and BAO data sets. In all five model-data set combinations, we have obtained the highest value of $H_0$ for the CC, PN$^{+}$\& SH0ES combination. This observation also align with the elevated $H_0$ reported by the SH0ES team (R22), specifically $H_0 = 73.30 \pm 1.04\, \text{km s}^{-1} \, \text{Mpc}^{-1}$ \cite{Riess_2022panplus}. The higher value is due to the inclusion of the SH0ES data points. We also noticed one think, when $H_0$ value increase then $\Omega_{m0}$ decrease and conversely $\Omega_{m0}$ increase then $H_0$ decrease.. This pattern shows that the major role of accelerating expansion is dark energy. For the power-law model, the models are largely consistent with each other at a high level of statistical confidence. This is also true for the other models. On the other hand, the hyperbolic and axion-like potentials yield wider fits for these parameters while also providing good fits to the model-specific parameters. 

The five models show good performance in comparison with $\Lambda$ as evidenced by the statistical metrics under consideration. These show promising possibilities for the models. The next phase of this analysis would be to consider the effect of the perturbative sector in comparison with observational data and how large-scale structure formation would be impacted. We intend to do this in future work, which will provide a more robust analysis of the models.
\section*{Appendix } \label{LCDMmodelappendix}
\begin{figure}[htbp]
 \centering
 \includegraphics[width=85mm]{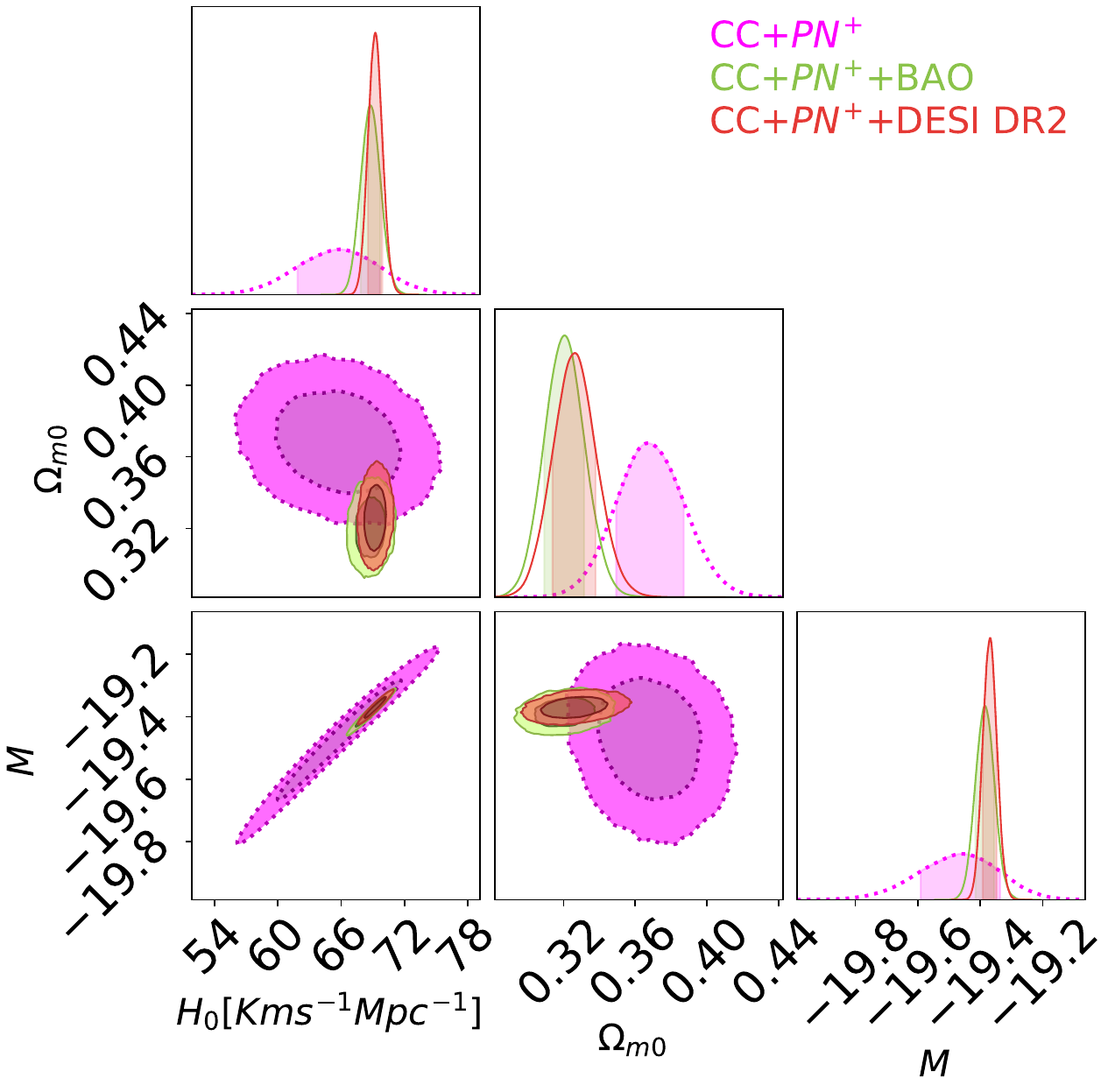}
 \includegraphics[width=85mm]{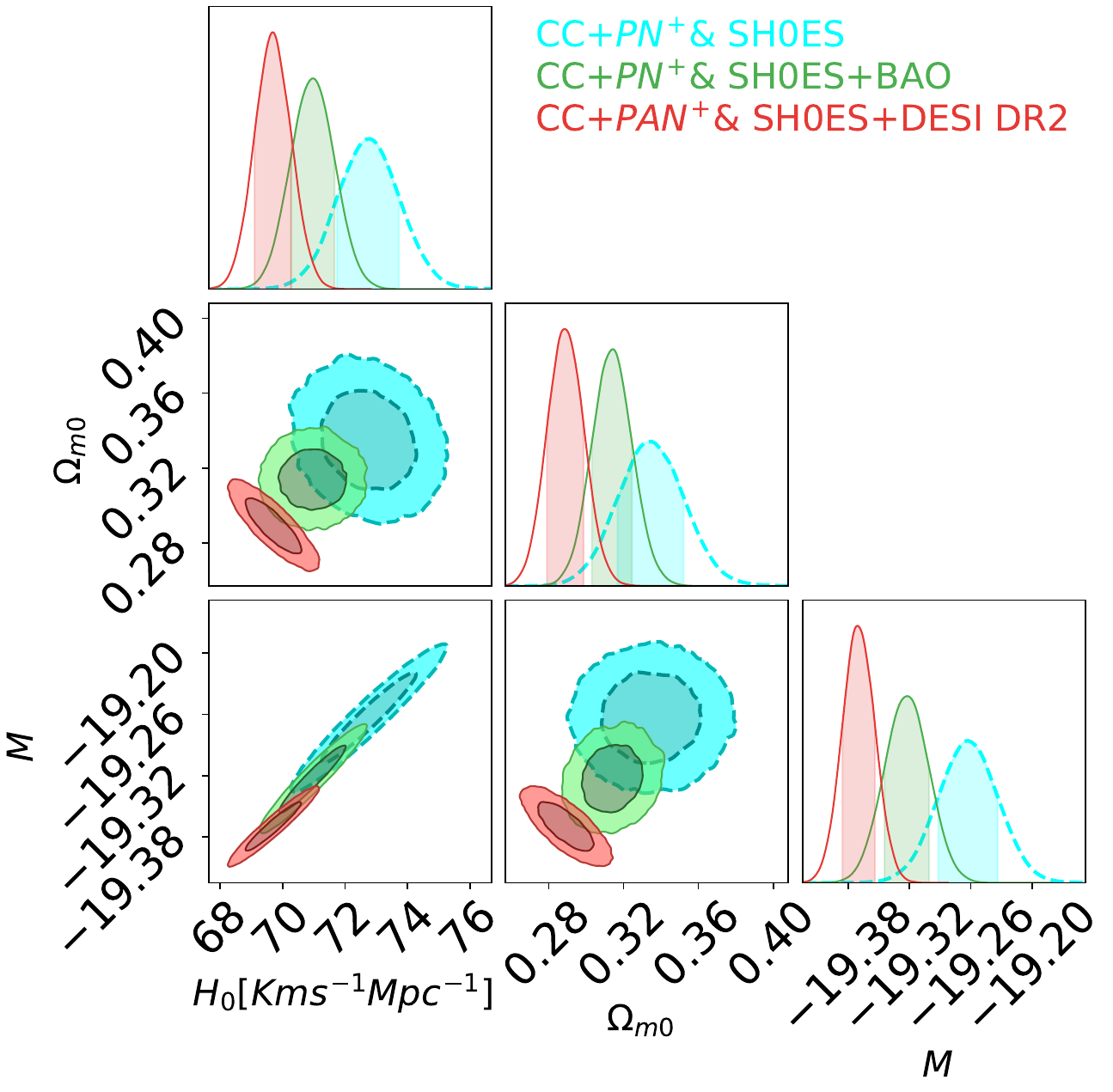}
 \caption{The marginalized posterior distributions and the corresponding 68\% and 95\% confidence contours for the $\Lambda$CDM model parameters, derived from the different dataset combinations.} \label{FigCDMMCMC}
 \end{figure}

The results of the $\Lambda$CDM model are presented in this section. All five potential functions have also been compared with the standard $\Lambda$CDM model. A visualization of the MCMC posteriors and confidence regions for the different data set combinations can be seen in Fig.~\ref{FigCDMMCMC}. Table-\ref{LCDM_outputs} presents the results for different data set combinations with the AIC and BIC statistical terms.

\begin{table}
\renewcommand{\arraystretch}{2}
    \centering
    \caption{The table presents results for the $\Lambda$CDM  model, with the first column listing the data set combinations. The second column shows the constraints on the Hubble constant $H_0$, while the third and fourth columns present the values for the matter density $\Omega_{m0}$ and nuisance parameter $M$ respectively. The fifth, sixth and seventh columns display the statistical terms $\chi^2_{min.}$, AIC and BIC respectively.}
    \label{LCDM_outputs}
    \begin{tabular}{ccccccc}
        \hline
		Data sets & $H_0\,\text{km} \, \text{s}^{-1} \text{Mpc}^{-1}$ & $\Omega_{m0}$ & $M$&$\chi^2_{min}$ & AIC & BIC\\ 
		\hline
		CC+$PN^{+}$ & $66.0^{+3.7}_{-4.1}$ & $0.367^{+0.020}_{-0.017}$ & $-19.46^{+0.12}_{-0.14}$ &1792.30 & 1798.30 & 1801.80  \\
		CC+$PN^{+}$+BAO & $68.70^{+1.01}_{-0.87}$ & $0.320\pm 0.011$ & $-19.385^{+0.030}_{-0.031}$ &1817.75 & 1823.75 & 1827.36\\ 
		CC+$PN^{+}$+DESI DR2 & $69.24^{+0.65}_{-0.71}$ & $0.327^{+0.011}_{-0.014}$ & $-19.368^{+0.020}_{-0.025}$ &1815.04 &1821.04 & 1824.64 \\ 
          \cline{1-7}
		CC+$PN^{+}\&$ SH0ES & $72.76^{+0.96}_{-1.03}$ & $0.335^{+0.017}_{-0.018}$ & $-19.264^{+0.031}_{-0.028}$&1539.22  &1545.22 & 1548.93 \\ 
		CC+$PN^{+}\&$ SH0ES+BAO &$70.98^{+0.68}_{-0.72}$ & $0.314^{+0.010}_{-0.011}$ & $-19.322^{+0.021}_{-0.022}$ & 1567.17 & 1573.17 & 1576.90 \\ 
		CC+$PN^{+}\&$ SH0ES+DESI DR2 & $69.68\pm 0.59$ & $0.2880^{+0.0106}_{-0.0089}$ & $-19.371^{+0.017}_{-0.015}$& 1575.5 & 1581.50 & 1585.22\\ 
		\hline
    \end{tabular}
\end{table}

\section*{Acknowledgements} JLS would like to acknowledge funding from Cosmology@MALTA which is supported by the University of Malta. BM acknowledges ANRF for the Mathematical Research Impact Centric Support (MATRICS)[File No : MTR/2023/000371]. The authors would like to acknowledge support from the Malta Digital Innovation Authority through the IntelliVerse grant.  This paper is based upon work from COST Action CA21136 {\it Addressing observational tensions in cosmology with systematics and fundamental physics} (CosmoVerse) supported by COST (European Cooperation in Science and Technology). We acknowledge the computing cluster Pegasus of IUCAA, Pune, India for using their computational facilities. 

\end{document}